\documentclass[a4paper,11pt]{article}

\usepackage{colortbl}
\usepackage{longtable}

\usepackage{textcomp}
\usepackage{amssymb}
\usepackage{units}

\usepackage[overload]{textcase}

\usepackage{graphicx}
\usepackage{subfigure}

\usepackage[labelsep=period]{caption}

\usepackage{multicol}

\usepackage{amssymb}

\usepackage{amsmath}
\usepackage{mathabx}

\usepackage[sort&compress]{natbib}
\bibpunct{[}{]}{,}{n}{}{}

\usepackage{sectsty}
\sectionfont{\centering\MakeTextUppercase}
\subsectionfont{\nohang\centering}
\subsubsectionfont{\nohang\centering\emph}

\makeatletter
\def\@seccntformat#1{\csname the#1\endcsname.\quad}
\makeatother

\begin{document}

\begin{titlepage}
\thispagestyle{plain}
\setcounter{page}{1}

\vfill

\begin{center}

   \vspace{0.3cm}
   {\Large \textbf{Measurement of Newtonian fluid slip using a torsional ultrasonic oscillator}} \\
   \vspace{0.3cm}
   {\large G. R. Willmott$^{*\dag\ddag}$, J. L. Tallon$^*$}\\
   \vspace{0.3cm}
   {\large $^*$Present Address: Industrial Research Limited, 69 Gracefield Rd, Lower Hutt, New Zealand.}\\
	\vspace{0.3cm}
	{\large $^\dag$ Corresponding author.}\\
	\vspace{0.3cm}
	{\large $^\ddag$ Email: g.willmott@irl.cri.nz}\\
	\vspace{0.3cm}
    {\large Phone: (64) (0)4 931 3220}\\
    \vspace{0.3cm}
    {\large Fax: (64) (0)4 931 3117}\\

\renewcommand{\abstractname}{}

\begin{abstract}

\noindent 
The composite torsional ultrasonic oscillator, a versatile experimental system, can be used to investigate slip of Newtonian fluid at a smooth surface. A rigorous analysis of slip-dependent damping for the oscillator is presented. Initially, the phenomenon of finite surface slip and the slip length are considered for a half-space of Newtonian fluid in contact with a smooth, oscillating solid surface. Definitions are revisited and clarified in light of inconsistencies in the literature. We point out that, in general oscillating flows, Navier's slip length $b$ is a complex number. An intuitive velocity discontinuity parameter of unrestricted phase is used to describe the effect of slip on measurement of viscous shear damping. The analysis is applied to the composite oscillator and preliminary experimental work for a 40~kHz oscillator is presented. The Non-Slip Boundary Condition (NSBC) has been verified for a hydrophobic surface in water to within $\sim 60$~nm of $\lvert b\rvert=0$~nm. Experiments were carried out at shear rate amplitudes between 230 and 6800~s$^{-1}$, corresponding to linear displacement amplitudes between 3.2 and 96~nm. 
\end{abstract}
\end{center}

\vspace{0.3cm}
\noindent{\large Receipt Date: 6 September 2007}\\
\vspace{0.3cm}
   \vspace{0.1cm}

\vfill

\end{titlepage}

\setcounter{page}{2}

\renewcommand{\thesection}{\arabic{section}}
\renewcommand{\thesubsection}{\Alph{subsection}}
\renewcommand{\thesubsubsection}{\arabic{subsubsection}}
 
\setcounter{section}{0}

\begin{center}
\section{Introduction}
\end{center}

In fluid mechanics, the Non-Slip Boundary Condition (NSBC) states that fluid at the interface between a fluid and a solid surface is stationary with respect to that surface. Surface slip arises when there is non-zero relative motion between the interfacial fluid and the solid surface, in which case the NSBC does not apply. In the vast majority of experiments described historically, the NSBC has been an adequate boundary condition \cite{303, 335}. However, finite surface slip has been the subject of much recent interest and promises to play a significant role in the current and future development of micro- and nanofluidic devices \cite{303, 321, 502, 335, 325, 403}. Navier \cite{400} first described surface slip as early as the 1820s and his mathematical definition of the \textquoteleft slip length' has been widely used, for example in recent reviews \cite{303, 321, 502, 335, 325} and analyses of slip-dependent flow \cite{305, 404, 329, 411}.

Slip is a well-studied phenomenon in non-Newtonian fluids such as polymer solutions \cite{335, 336, 411, 433}. For slip of Newtonian fluids (Newtonian slip), theory is well developed \cite{502}, but interpretation of experiments has been much more controversial. Newtonian slip is important because technologies can usefully be applied to Newtonian fluids, but also because understanding of Newtonian slip will provide a more profound understanding of fundamental slip mechanics. The physical mechanisms governing slip are affected by variables that include surface chemistry, shear rate, surface structure, Newtonian viscosity, non-Newtonian behaviour at the molecular scale and combinations of each of these variables \cite{335}. Nanoengineered \textquoteleft superhydrophobic' surfaces trap gas between the fluid and the surface, producing large slip lengths of the order $\sim$10 $\mu$m \cite{430, 437, 320, 436, 434, 435}.

Experimentally, Newtonian slip has been investigated using a variety of methods \cite{303}. The most widely used and precise technique involves measurement of the drainage forces between closely-spaced surfaces in an Atomic Force Microscope (AFM) or Surface Force Apparatus (SFA) and applying an analysis based on Reynolds' theory of lubrication \cite{305, 415}. Robust, non-zero values of $b$ for Newtonian fluids at relatively smooth surfaces are typically of the order 10-100~nm \cite{304, 324, 393, 427, 318, 325, 334, 439, 441}. Experiments have not established firm quantitative correlations between slip and major experimental variables, such as contact angle \cite{303, 335}. Recently, it has emerged with some clarity that in AFM or SFA experiments, the NSBC can be expected to hold on smooth hydrophilic surfaces \cite{393, 427, 413, 410}, while finite slip lengths of order 10-20~nm have been measured on some hydrophobic surfaces \cite{393, 427}. Additionally, the long-held idea that surface roughness should inhibit slip \cite{335, 483} is now supported by reasonable experimental evidence \cite{335, 393, 427}.

Shear-mode oscillations can be used to drive fluid flow, and such systems pose scientific questions that are distinct from those relating to the study of steady fluid flows. Oscillating piezoelectric components are likely to prove very useful in future applications. Ultrasonic oscillators, well established as important tools for investigating thin films and bulk fluids, can also be used to probe the interfacial forces that are relevant in the study of slip. Kanazawa and Gordon \cite{331} analytically coupled a half-space of Newtonian fluid to a piezoelectric Quartz Crystal Microbalance (QCM), in which a solid interface oscillates in shear mode, using the NSBC. Further reports have presented experiments \cite{306, 313, 319, 316} and analysis \cite{319, 313, 306, 308, 309, 314} describing slip of a Newtonian fluid adjacent to an oscillating surface. A relatively large set of slip lengths measured using a QCM was obtained by Ellis and Hayward \cite{306}. 

Torsional quartz crystal oscillators operating at kHz frequencies were initially developed and used for viscoelasticity measurements \cite{432} and have been used to measure the complex and viscoelastic behaviour of polymer solutions. More recent work has investigated dispersions of colloidal \cite{425, 426, 431} and micellar \cite{419} spheres. The shear rates accessible using torsional oscillators are considered advantageously high for such rheological measurements \cite{419, 431}. A particular configuration, the composite torsional ultrasonic oscillator, was developed and analysed by Robinson et al. \cite{301} and Robinson and Smedley \cite{302}, who also derived an equivalent circuit. Recently, the first experimental results investigating Newtonian slip at a smooth surface using such a device have been reported \cite{418}, for oscillations at 40~kHz.   

There has been some confusion relating to fundamental aspects of slip, perhaps best characterised by Neto et al.~\cite{303}, who noted that \textquotedblleft dialogue between the QCM community and other researchers interested in slip has been scarce\textquotedblright. In order to address this issue, Navier's definition of the slip length is used consistently in the current work. An intuitive, consistent and general methodology for slip measurement at an oscillating surface is presented. Analytic inconsistencies in the literature are identified and clarified, especially those concerning an oscillating interface. The composite torsional oscillator is introduced along with a full analysis for measurement of Newtonian slip using this device. Preliminary experiments are presented to demonstrate how this device is used to probe interfacial forces in oscillatory flow. We discuss the interpretation of measurements and the analytical assumptions on which they are based. Finally, methods of slip measurement are compared and the advantages of using the composite torsional oscillator are explained. 

\section{\label{Theory}Theory}

\subsection{\label{Gen slip}Newtonian slip at a smooth, oscillating surface}
	
\subsubsection{\label{Newtonian Flows}Flow profile for shear mode oscillations}

By definition, viscosity $\eta$ is constant and independent of shear rate throughout a Newtonian fluid. The velocity field $\mathbf{u}$ of an incompressible Newtonian fluid of density $\rho$ in the presence of pressure gradient $\nabla\mathbf{p}$ and body force $\mathbf{F}$ can be determined by solving the Navier-Stokes equation,  

\begin{equation}\label{eq:Navier-Stokes}
\begin{split}
\rho\left(\frac{\partial \mathbf{u}}{\partial t}+\mathbf{u}.\nabla \mathbf{u}\right)=\mathbf{F}-\nabla\mathbf{p}+\eta\nabla^2\mathbf{u},\\
\nabla.\mathbf{u}=0.
\end{split}
\end{equation}

\noindent Microfluidic flows have characteristically low Reynolds numbers, so the convective derivative $\rho\mathbf{u}.\nabla\mathbf{u}$ is insignificant when compared with the viscous term $\eta\nabla^2\mathbf{u}$. Solutions to the Navier-Stokes equation are dependent on flow geometry and boundary conditions. 

The relevant solution for the current paper concerns a smooth surface oscillating in shear mode in contact with a fluid half-space. In the absence of significant body forces or pressure differentials, the equation of the shear wave induced in the fluid is

\begin{equation}\label{eq:Gen fluid wave}
u_f\left(z,t\right)=u_{f0}\left(t\right)\exp\left(-\frac{\sqrt{2i}}{\delta}z\right),
\end{equation}

\noindent where $u_f$ is the fluid velocity in the direction of the surface oscillations, $z$ is the Cartesian direction perpendicular to the solid surface and $\delta=\left(\frac{2\eta}{\omega\rho}\right)^{\frac{1}{2}}$. The value of $u_f$ at the interface is $u_{f0}$, which depends on the fluid-surface boundary condition. The time dependence of $u_f$ and $u_{f0}$ is $e^{i\omega t}$, where the frequency of oscillation is $\omega$ and $t$ is time. This solution is relevant for ultrasonic oscillators.


\subsubsection{\label{Nsl}Navier's slip length}

Boundary slip occurs when the NSBC is contravened, so that there is finite relative motion between a solid surface and an adjacent fluid. Navier \cite{400} characterised the slip boundary condition by defining the slip length $b$. For flow adjacent to a planar, impermeable solid surface, 

\begin{equation}\label{eq:Gen slip length}
u_{\parallel 0}\left(t\right)=b\frac{\partial u_{\parallel}\left(z,t\right)}{\partial z}\vline_{z=0}.
\end{equation}

\noindent Here $u_{\parallel}$ is the shear velocity relative to and parallel with the surface and $u_{\parallel 0}$ is the value of $u_{\parallel}$ at $z=0$. Navier's definition of slip has been widely used over a long period and has a strong physical basis relating to viscous force at the interface. The fluid shear rate at the interface that appears in Eq.~\ref{eq:Gen slip length} also determines the viscous force,

\begin{equation}\label{eq:Gen viscous stress}
\sigma_{xz}\left(t\right)=\eta\frac{\partial u_{\parallel}\left(z,t\right)}{\partial
z}\vline_{z=0}.
\end{equation}

\noindent Slip therefore determines the shear stress at the interface, which is analogous to friction in character. The coefficient $k$, defined as 

\begin{equation}\label{eq:friction define}
\sigma_{xz}\left(t\right)=k u_{\parallel 0}\left(t\right),
\end{equation}

\noindent is finite only when there is slip. Using Eqs.~\ref{eq:Gen slip length}, \ref{eq:Gen viscous stress} and \ref{eq:friction define}, we find

\begin{equation}\label{eq:b friction}
b=\frac{\eta}{k}.
\end{equation}

\noindent It is noted that the NSBC and boundary slip are, respectively, analogous to the regimes of \textquoteleft static' and \textquoteleft sliding' solid-solid friction. Extending the analogy, we might expect a wide variety of material and interfacial properties to determine the onset and magnitude of solid-liquid friction. If slip length depends on shear rate, Eq.~\ref{eq:Gen slip length} becomes non-linear and rate-dependent formulations of slip (e.g. \cite{482}) become relevant. Currently, there is conflicting experimental evidence relating to possible shear rate dependence of $b$ \cite{303, 335}. 

\begin{figure}
\begin{center}
\includegraphics[width=8.5cm]{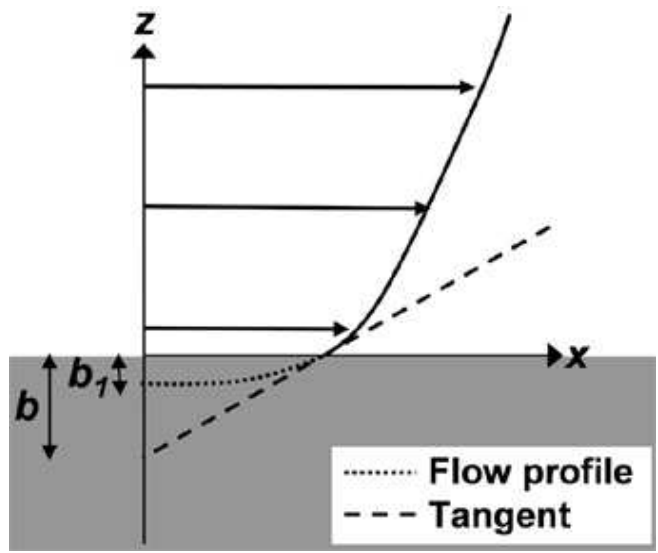}
\end{center}
\caption{
The slip lengths $b$ and $b_1$ defined in the text, with the shaded area representing solid material. Navier's slip length $b$ (Eq.~\ref{eq:Gen slip length}) is only equal to the variable $b_1$ (Eq.~\ref{eq:b_1}) when the velocity profile is linear. The labels $x$ and $z$ denote Cartesian coordinate axes.} 
\label{Fig:Slip definition}
\end{figure}

A different physical interpretation of slip length relates $b$ to the distance into the solid over which the tangent to the flow profile at the surface must be extrapolated for the velocity to reach zero. In this context, Navier's slip length has correctly been described as a fictional distance \cite{327, 335} and is labelled in Fig.~\ref{Fig:Slip definition}. The definition of slip length used in some recent slip studies \cite{306, 308, 309, 314}, and formalised by Ellis and Hayward \cite{306} as (their notation) $v_P(d_P)=v_B(d_P-b)$, or

\begin{equation}\label{eq:b_1}
u_{\parallel}\left(-b_1,t\right)=0,
\end{equation}

\noindent is distinct from Navier's definition in Eq.~\ref{eq:Gen slip length}. Ellis and Hayward's definition corresponds to extrapolation of the flow profile a distance $b_1$ (Fig.~\ref{Fig:Slip definition}) relative to the surface when slip occurs. The value of $b_1$ is equal to $b$ only in the case of a linear velocity flow profile. An often-used schematic diagram \cite{303, 306, 411, 335, 308, 404, 444}, which uses a linear flow profile to depict slip length, appears to have been an important factor in the emergence of Eq.~\ref{eq:b_1}.

\subsubsection{An intuitive slip parameter}

We wish to point out that when the flow of a Newtonian fluid is driven by an oscillating surface (as introduced above), the shear rate introduces non-zero phase into the value of $b$, which is defined using $u_{\parallel}$ (Eq.~\ref{eq:Gen slip length}). The existence of this phase highlights the limitations of direct quantitative comparison of measurements of $b$ using different experimental techniques.

When measuring slip at an oscillating surface, it is both intuitive and useful to use the dimensionless complex parameter $\alpha=\lvert\alpha\rvert e^{i\phi_\alpha}$. This parameter, which simply quantifies the velocity discontinuity between the solid surface and the first liquid layer, simplifies the algebra describing the effect of damping on the equivalent circuit of a piezoelectric oscillator.  
It is defined by interfacial properties and does not depend on the flow profile,  

\begin{equation}\label{eq:Gen alpha define}
\alpha u_s\left(t\right)=u_{f0}\left(t\right).
\end{equation}

\noindent The velocity of the solid surface is $u_s$ and the fluid oscillation is assumed to be dominated by a sinusoidal oscillation of angular frequency equal to that of the solid surface. Under this assumption, $\alpha$ is equivalent to parameters introduced by Ferrante et al. \cite{319} and used elsewhere \cite{312, 313} for modelling of slip at an oscillating surface; to our knowledge, it has not previously been applied to experimental measurements with full consideration of phase dependence. More generally, $\alpha$ can be decomposed into Fourier components, an approach that has been used when modelling nanobubble-induced slip \cite{307, 315}. The NSBC is consistent with the values $\lvert\alpha\rvert=1$ and $\phi_\alpha=0$.

For an oscillating surface, flow is relative to surface motion, so 

\begin{equation}\label{eq:Gen osc flow profile}
\begin{split}
u_{\parallel}\left(t\right)&=(u_f\left(t\right)-u_s\left(t\right))\\
&=u_s\left(t\right)\left(\alpha\exp\left(-\frac{\sqrt{2i}}{\delta}z\right)-1 \right).
\end{split}
\end{equation}

\begin{figure}
\begin{center}
\subfigure[]{\label{Fig:c}\includegraphics[width=8.5cm]{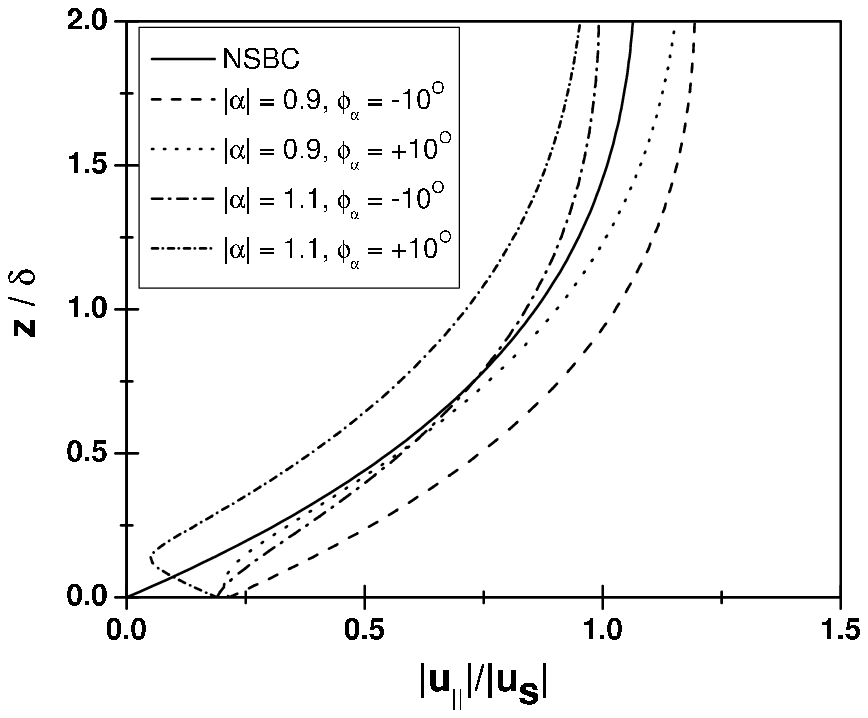}}
\subfigure[]{\label{Fig:d}\includegraphics[width=8.5cm]{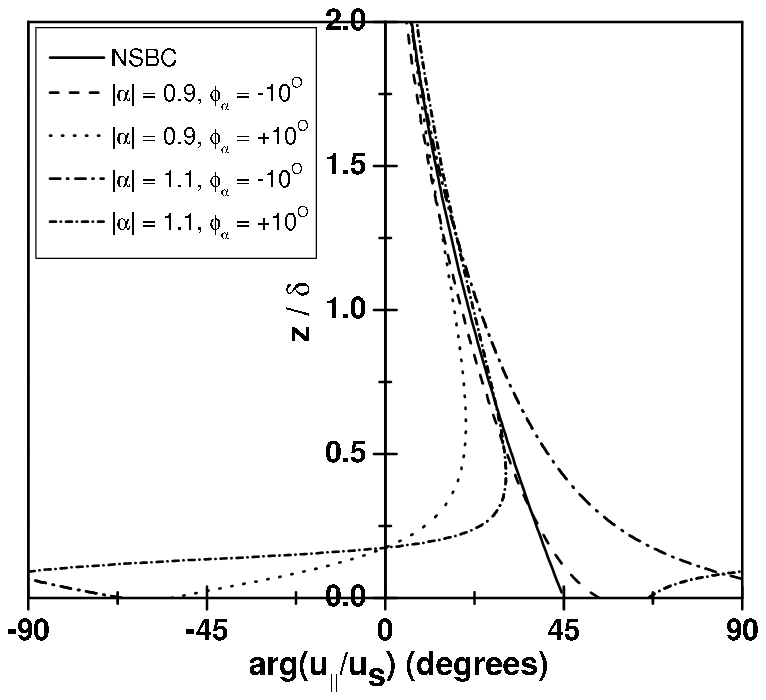}}
\end{center}
\caption{Magnitude (Fig.~\ref{Fig:c}) and phase (Fig.~\ref{Fig:d}) of $u_\parallel$/$u_s$ as a function of the distance from an oscillating surface for a Newtonian fluid, calculated using Eq.~\ref{eq:Gen osc flow profile} and plotted for the NSBC as well as various slip boundary conditions. The NSBC phase is discontinuous at the interface, taking a value of zero at $z=0$ even though the limit as $z$/$\delta$ approaches zero is $45^\degree$.
} \label{Fig:Slip phasor}
\end{figure}

\begin{figure}
\begin{center}
\subfigure[]{\label{Fig:alpha_b_amp}\includegraphics[width=8.5cm]{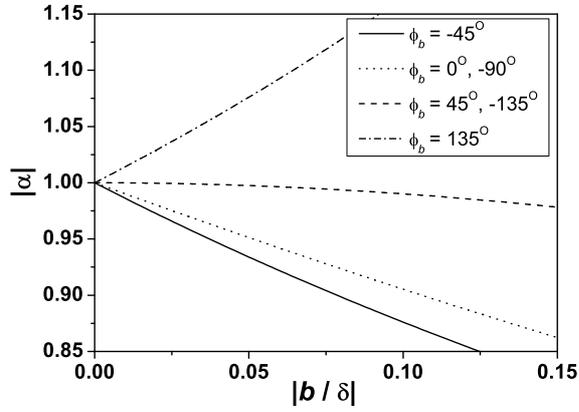}}
\subfigure[]{\label{Fig:alpha_b_phase}\includegraphics[width=8.5cm]{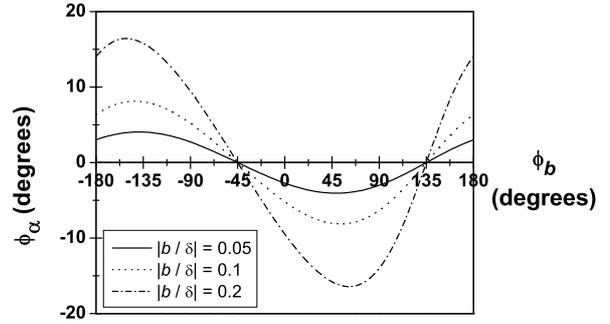}}
\end{center}
\caption{Magnitude (Fig. \ref{Fig:alpha_b_amp}) and phase (Fig. \ref{Fig:alpha_b_phase}) of $\alpha$ relative to $b$ for a Newtonian fluid, calculated using Eq.~\ref{eq:Gen alpha equals}. Equations \ref{eq:Gen slip length} and \ref{eq:Gen osc flow profile} can be used to show that $\phi_b$ takes a value of $-45^\degree$ rather than $135^\degree$ when $\phi_\alpha=0$. A value of $\phi_b=135^\degree$ is consistent with a negative value of $b$ and NSBC phase dependence.} \label{Fig:alpha vs b}
\end{figure}

\noindent The magnitude and phase of $u_\parallel$ relative to $u_s$ are plotted in Fig.~\ref{Fig:Slip phasor}, for both slip and non-slip cases. Figure~\ref{Fig:c} shows that the relative amplitude of oscillation is uniformly non-zero at the surface when slip is present, but varies significantly near the interface ($\frac{z}{\delta}\lesssim 0.25$) depending on the values of $\lvert\alpha\rvert$ and  $\phi_\alpha$. Further from the surface ($\frac{z}{\delta}\gtrsim 1$), the relative amplitude tends towards a stable value close to 1 but dependent on $\lvert\alpha\rvert$. The relative phase follows a similar trend in Fig.~\ref{Fig:d}, showing strong dependence on the combination of $\lvert\alpha\rvert$ and $\phi_\alpha$ near the interface. There is a phase discontinuity at $\frac{z}{\delta}\approx 0.1$ for the cases in which $\lvert\alpha\rvert > 1$. The plane of discontinuous phase is consistent with a negative slip length (Fig.~\ref{Fig:alpha vs b}) and could, for example, be interpreted as the edge of a stagnant boundary layer. Further from the surface ($\frac{z}{\delta}\gtrsim 1$), the relative phase converges towards the NSBC value, which tends towards zero a long way from the interface.

Using this flow profile, slip length can be directly related to $\alpha$ (Eqs.~\ref{eq:Gen fluid wave}, \ref{eq:Gen slip length}, \ref{eq:Gen alpha define} and \ref{eq:Gen osc flow profile}):

\begin{equation}\label{eq:Gen alpha equals}
\alpha=\frac{1}{1+\frac{\sqrt{2i}}{\delta}b}.
\end{equation}

\noindent Equation \ref{eq:Gen alpha equals} is plotted in Fig. \ref{Fig:alpha vs b} using the notation $b=\lvert b\rvert e^{i\phi_b}$. Figure \ref{Fig:alpha_b_amp} shows that $\lvert\alpha\rvert$ decreases when $\phi_b$ is within $90^\degree$ of $-45^\degree$, so that the slip length $b$ is positive. The phase plot (Fig. \ref{Fig:alpha_b_phase}) demonstrates that $\phi_b=-45^\degree$ when the fluid oscillates in phase with the surface ($\phi_\alpha=0^\degree$). For the NSBC, $b=0$, so $\phi_b$ can take any value. The maximum size of the phase shift $\phi_\alpha$ is limited by slip length magnitude. The complicated nature of the flow profile near the interface supports the use of the parameter $\alpha$, which does not depend on the flow other than at the interface itself, in contrast to $b$. However, calculation of the shear rate using the flow profile is still necessary for derivation of $\alpha$ from surface force experiments. 

\subsubsection{Effect of slip on damping force and mechanical impedance}

The damping force on an oscillating surface is equal and opposite to the viscous force in the adjacent Newtonian fluid. When slip occurs, the damping stress is (Eqs.~\ref{eq:Gen fluid wave}, \ref{eq:Gen viscous stress} and \ref{eq:Gen osc flow profile}),   

\begin{equation}\label{eq:Gen interface stress}
\begin{split}
\sigma_{xz}\left(t\right)=-\frac{\sqrt{2i}}{\delta}\eta \alpha u_s\left(t\right).
\end{split}
\end{equation}

Analysis of ultrasonic oscillators is greatly simplified by using the equivalent circuit analogy, in which mechanical components are replaced by their electronic equivalents. We now incorporate the damping stress into the analysis of a mechanical circuit. The stress at the interface (Eq.~\ref{eq:Gen interface stress}), which has components both in phase and out of phase with surface velocity, can be written as 

\begin{equation}\label{eq:Components of interface stress}
\begin{split}
\sigma_{xz}\left(t\right)=-\left(R+iX\right)u_s\left(t\right).
\end{split}
\end{equation}

\noindent The impedance of the system is $Z=R+iX$, as previously applied to a torsional oscillator by Bergenholtz et al. \cite{425}, who used the NSBC. This impedance is dependent on the liquid, the surface, and the interaction between them. The real part of the impedance ($R$) determines the stress opposing, and in phase with, the surface velocity,

\begin{equation}\label{eq:R stress}
R=\frac{\sqrt{2}\eta\lvert\alpha\rvert}{\delta}\cos\left(\frac{\pi}{4}+\phi_\alpha\right).
\end{equation}

\noindent This impedance is viscous in nature and dissipative of energy. In an equivalent circuit, velocity is analogous to current and $R$ is like a resistance. Energy dissipation is determined by the quality factor of the oscillation.

The inertial term ($X$) gives the stress that is in phase with the acceleration of the surface,

\begin{equation}\label{eq:X stress}
X=\frac{\sqrt{2}\eta\lvert\alpha\rvert}{\delta}\sin\left(\frac{\pi}{4}+\phi_\alpha\right).
\end{equation}

\noindent An increase in the inertial term is equivalent to the addition of mass to the surface of the oscillator. In the equivalent circuit, the time derivative of current is analogous to acceleration (modified by a factor of $\omega$), so $X$ is an inductance. A change in $X$ is measured by considering the change in the period of oscillation. Under the NSBC, $R$ and $X$ are equal for Newtonian fluid damping. Slip-induced change in the relative sizes of these terms indicates a finite value of $\phi_\alpha$.

\subsubsection{\label{Ci}Conceptual issues}

The theory that has been presented implicitly addresses two conceptual pitfalls that have caused confusion and misinterpretation in a range of previous work. The first pitfall concerns the variant \textquoteleft slip length' $b_1$, defined in Eq.~\ref{eq:b_1}, which is only equal to $b$ 
in the limit of a linear flow profile. 
This limit is generally not applicable for slip measurements. The difference between $b$ and $b_1$ is greater than 10\% for typical slip measurements using AFM, SFA or ultrasonic oscillation methods. We further note that Ellis and Hayward's definition has been used inconsistently; both with and without considering a Taylor expansion \cite{309, 314}. Navier's slip length $b$ should be used at the present time due to its strong physical basis and to promote consistency across experimental studies.




The second conceptual pitfall relates to those experimental studies of slip at oscillating surfaces which have imposed a restriction on the phase of slip \cite{306, 309, 314, 308}, especially the assumption that the slip length $b$ is real-valued. It is equivalent to assume that $k$ or its reciprocal $s$ is real-valued, or more generally, that the damping force is in phase with the surface velocity (Eq.~\ref{eq:b friction}). Several lines of reasoning lead us to emphasize that $b$ must be allowed to have non-zero phase. We have shown that, because the slip length is defined using the shear rate (Eq.~\ref{eq:Gen slip length}), the $z$-dependence of the fluid velocity (a complex exponential, Eq.~\ref{eq:Gen fluid wave}) introduces phase to the value of $b$. The usual (NSBC) form of solid-fluid friction force is viscous (Eq.~\ref{eq:Gen interface stress}) and therefore $45^\degree$ out of phase with the velocity above an oscillating surface. Figure~\ref{Fig:alpha vs b} shows that the phase of $b$ in the limit approaching the NSBC is non-zero and that both magnitude and phase must be determined for either $b$ or $\alpha$ to be fully described. Restricting $b$ to real values is equivalent to restricting $\alpha$ to one degree of freedom in Eq.~\ref{eq:Gen alpha equals}. Furthermore, experiments \cite{418} have indicated that a fluid can be expected to oscillate in phase with, or slightly lag, the surface oscillation. Neither interpretation is consistent with a real value of $b$. 

Having established that $b$ must be considered complex, we address specific arguments made in previous studies in support of a real-valued $b$. Firstly, there has been a direct assumption \cite{306, 309, 314} that the value of $k$ (or $s$) should not be phase dependent. This assumption has been based on Rodahl and Kasemo's argument \cite{395}, which links $k$ to the ratio of force to velocity for low Reynolds' number drag on small bodies (e.g. Stokes drag), in which case $k$ is a real constant. This argument ignores the fact that important factors in the determination of Stokes drag include the contribution of pressure surrounding the particle, the requirement for steady non-oscillatory flow and use of the NSBC. We note that friction and drag parameters are usually real because they are not used to describe oscillatory flow. In conjunction with this point, we note that the term \textquoteleft coefficient of friction', which has frequently been used 
to describe $k$ in slip studies, is distinct from the usual parameter of that name, which relates the shear force to the normal force in the case of solid-solid sliding friction. Secondly, an argument for phase restriction \cite{306} has drawn on the correlation between bond strength and bond length, which should determine magnitude and phase of slip respectively. Such an argument does not explain why the phase should take one particular value, or why such a correlation between magnitude and phase should not be experimentally investigated.

Non-zero phase is not significant in previous work relating to lower-frequency oscillations in AFM or SFA devices \cite{393}, in which $\delta\sim$100~$\mu$m for 39~Hz oscillations in water, as opposed to $\sim$500~nm for a 1~MHz oscillator. Similarly, periodically patterned microchannels \cite{317} are adequately described by a real-valued slip length when the ratio of the flow velocity to the pattern wavelength does not exceed $\sim$1~kHz, corresponding to patterns of extremely short spatial wavelength.

\subsection{\label{T.O.}Slip measurement using the composite torsional ultrasonic oscillator}

\subsubsection{\label{Theory and TO}Theory applied to the torsional oscillator}

\begin{figure}
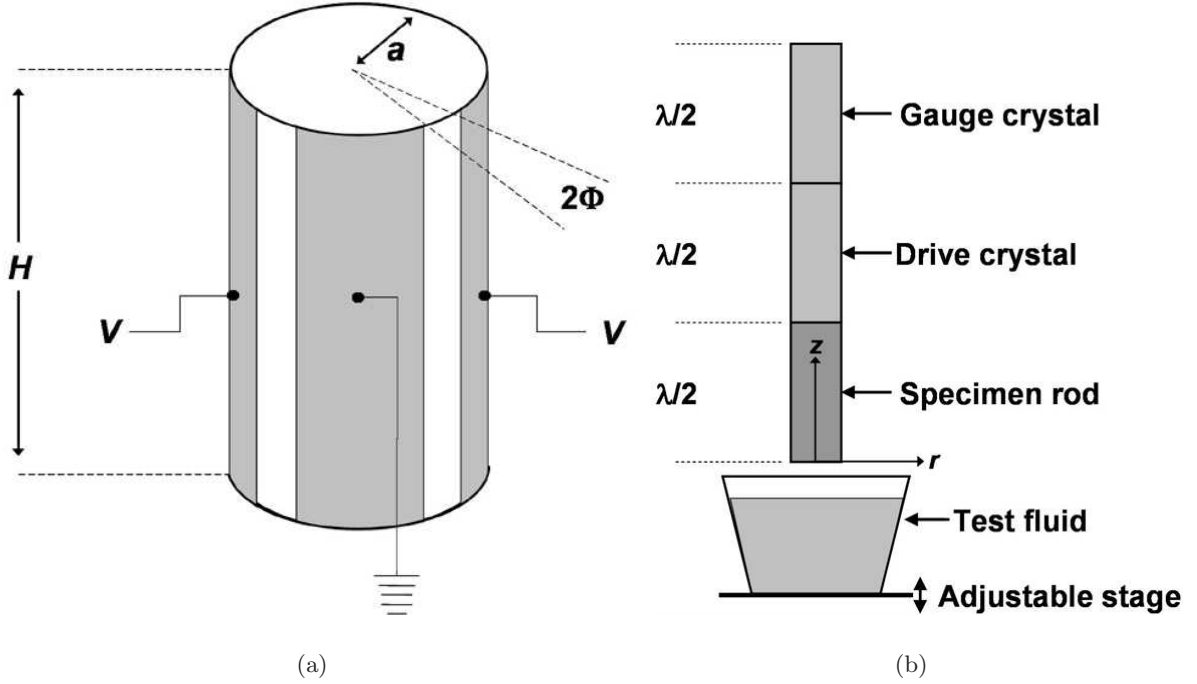

\begin{center}
\subfigure[]{\label{Fig:e}\includegraphics[width=8.5cm]{TO_Schematic_onerod}}
\subfigure[]{\label{Fig:f}\includegraphics[width=7cm]{TO_Schematic_III}}
\end{center}
\caption{The composite torsional ultrasonic oscillator \cite{301,302}. Figure~\ref{Fig:e} shows a piezoelectric $x$-cut $\alpha$-quartz rod, which takes dimensions of height $H\approx 48$mm and of radius $a\approx 5$~mm for the rods of resonant frequency 40~kHz used in experiments presented here. Four conducting electrodes, separated from each other by an angle $2\Phi$, are coated on to the rod surface and a voltage $V$ is applied in order to excite torsional oscillations. In the composite oscillator (Fig.~\ref{Fig:f}) two identical piezoelectric rods are joined together with a co-resonant specimen rod. A standing torsional wave of wavelength $\lambda$ is set up with antinodes located at free ends and joints. Typically the length of each rod is $\lambda$/2, as in the figure. The apparatus is held at the stationary nodes, where the electrode connections are also located. Viscous fluid damping is measured by raising the test fluid on an adjustable stage so that the specimen rod is immersed from the free end along a length of the rod $l$. The $z$- and $r$-axes of the cylindrical polar coordinate system used in the text are as shown. The voltages across the drive ($V_d$) and gauge ($V_g$) crystals are measured along with the resonant period $\tau$. 
} \label{Fig:TO_Schematic}
\end{figure}

The previous discussion of slip is now applied to the composite torsional ultrasonic oscillator (Fig. \ref{Fig:TO_Schematic}). The damping torque when the specimen rod is submerged in a Newtonian fluid was derived by Robinson et al. \cite{302} under the NSBC, using an equivalent circuit analysis. In that derivation, the shear damping stress on a flat surface is integrated over the surface of the specimen rod. Appendix~A, in this paper, gives the same derivation with slip included and Appendix~B discusses an assumption that is used. Similar derivations have considered damping of other torsional oscillators by viscoelastic fluids under the NSBC, yielding similar functional forms \cite{424, 425, 431}. For the oscillator of specimen rod radius $a\gg\delta$ described by Fig.~\ref{Fig:TO_Schematic}, the total torque for $h(=1$ or $2)$ immersed flat surfaces is

\begin{equation}\label{eq:TO NSBC T}
T\left(t\right)=a^2\left(\lambda+\frac{1}{2}h\pi a\right)\frac{\sqrt{2i}}{\delta}\eta\alpha u_{s,max}\left(t\right),
\end{equation}

\noindent where the subscript \textquoteleft max' refers to the value at the antinode of the standing wave, which has wavelength $\lambda$.

When a specimen rod is immersed, the resistance and inductance added to the equivalent circuit are $R_\eta$ and $L_\eta$ respectively, so  

\begin{equation}\label{eq:TO T equivalent}
T\left(t\right)=(R_\eta+i\omega L_\eta)\frac{u_{s,max}\left(t\right)}{a}.
\end{equation}

\noindent Equation \ref{eq:TO T equivalent} is directly analogous to Eq.~\ref{eq:Components of interface stress} for this specific configuration if we substitute notations $R=R_\eta$ and $X=\omega L_\eta$, consistent with the resistive and inductive descriptions of these parameters. 

Comparing Eqs.~\ref{eq:TO NSBC T} and \ref{eq:TO T equivalent} and introducing a constant $C$, we find

\begin{equation}\label{eq:R_eta}
\begin{split}
R_\eta &=a^3\left(\lambda+\frac{1}{2}h\pi a\right)\frac{\sqrt{2}\eta}{\delta}\lvert\alpha\rvert\cos\left(\frac{\pi}{4}+\phi_\alpha\right)\\
&=C\lvert\alpha\rvert\cos\left(\frac{\pi}{4}+\phi_\alpha\right)\\
L_\eta &=\frac{C}{\omega}\lvert\alpha\rvert\sin\left(\frac{\pi}{4}+\phi_\alpha\right)
\end{split}
\end{equation}

The equivalent circuit reduces to a series LCR circuit \cite{301}. $L_\eta$ is therefore related to the change in period upon immersion $\tau_\eta$ and the
resonant period $\tau$, using

\begin{equation}\label{eq:TO NSBC tor}
L_\eta=L_{tot}\frac{\tau_\eta}{\tau}.
\end{equation}

\noindent The inverse quality factor of the oscillation $Q^{-1}$ is given by the total resistance of the composite oscillator $R_{tot}$ and the total inductance $L_{tot}$:

\begin{equation}\label{eq:TO Q^-1 equivalent}
Q^{-1}= \frac{R_{tot}}{\omega L_{tot}}.
\end{equation}

\noindent $R_\eta$ is calculated by assuming that $L_{tot}$ is dominated by the mechanical inductance (moment of inertia) of the rods \cite{302}, so that ($L_\eta \ll L_{tot}$)

\begin{equation}\label{eq:TO R_eta}
R_\eta=\omega L_{tot}Q^{-1}_\eta.
\end{equation}

\noindent $Q^{-1}_\eta$ is the change in $Q^{-1}$ upon immersion. When piezoelectric elements are incorporated, the equivalent circuit for the torsional oscillator gives the inverse quality factor, 

\begin{equation}\label{eq:TO Q^-1}
Q^{-1}=K\left(\frac{V_d}{V_g}\right),
\end{equation}

\noindent where $V_d$ and $V_g$ are voltages applied to the drive and gauge crystals respectively, and $K$ is a calibrated constant for a particular oscillator arrangement \cite{301}. The quality factor is also given by

\begin{equation}\label{eq:TO Q^-1c}
Q^{-1}=\Bigl(\frac{\tau_1-\tau_2}{\sqrt{3}\tau}\Bigr),
\end{equation}

\noindent where $\tau$ is the resonant period and $\tau_1$ and $\tau_2$ are measurements of the period when $V_d$ is double the resonance value with $V_g$ held constant. By plotting the right hand side of Eq.~\ref{eq:TO Q^-1c} against the ratio $V_d$/$V_g$, Eq.~\ref{eq:TO Q^-1} can be checked and the value of $K$ determined. $K$ is independent of specimen rod damping, but it is dependent on the rigid moment of inertia of the specimen rod, so changes if mass is added to the specimen rod.

We now consider comparison of a base case (subscript 0) in which the NSBC holds ($\alpha=1$, $\phi_\alpha=0$), with a case in which slip is present (subscript 1). In practice, it is difficult to determine whether there is slip in the base case, so $\alpha$ can be used to measure the relative amplitude and phase of slip between two cases investigated. $C$ remains constant between measurements as long as $a$, $\lambda$, $\eta$, $\rho$ and $\omega$ are constant, so using Eqs.~\ref{eq:R_eta}, \ref{eq:TO NSBC tor} and \ref{eq:TO R_eta}, we define $\Delta Q^{-1}$ and $\Delta\tau$ as

\begin{equation}\label{eq:R_foralpha}
\begin{split}
\Delta Q^{-1}=\frac{Q^{-1}_{\eta 1}}{Q^{-1}_{\eta 0}}=\frac{R_{\eta 1}}{R_{\eta 0}}
=\sqrt{2}\lvert\alpha\rvert\cos\left(\frac{\pi}{4}+\phi_\alpha\right),
\end{split}
\end{equation}

\noindent and

\begin{equation}\label{eq:L_eta}
\begin{split}
\Delta\tau=\frac{\tau_{\eta 1}}{\tau_{\eta 0}}=\frac{L_{\eta 1}}{L_{\eta 0}}
=\sqrt{2}\lvert\alpha\rvert\sin\left(\frac{\pi}{4}+\phi_\alpha\right).
\end{split}
\end{equation}

Rearranging Eqs.~\ref{eq:R_foralpha} and \ref{eq:L_eta}, $\alpha$ is calculated directly from measurement:

\begin{equation}\label{eq:alpha measure mag}
\lvert\alpha\rvert=\sqrt{\frac{\left(\Delta Q^{-1}\right)^2}{2}+\frac{
\left(\Delta\tau\right)^2}{2}},
\end{equation}

\noindent and

\begin{equation}\label{eq:alpha measure phase}
\phi_\alpha=\arctan\left(\frac{\Delta \tau}{\Delta Q^{-1}}\right)-\frac{\pi}{4}.
\end{equation}

\begin{figure}
\begin{center}
\subfigure[]{\label{Fig:alpha amp}\includegraphics[width=8.5cm]{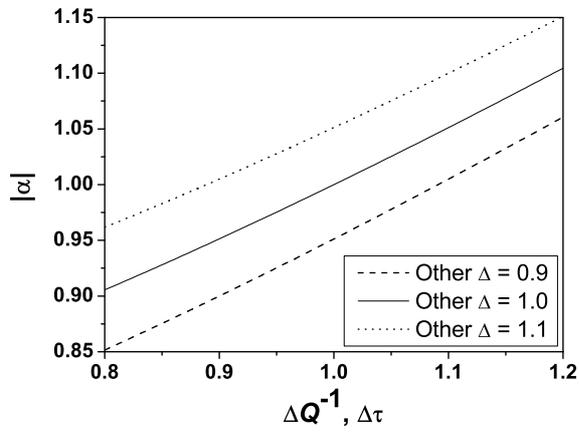}}
\subfigure[]{\label{Fig:alpha phase}\includegraphics[width=8.5cm]{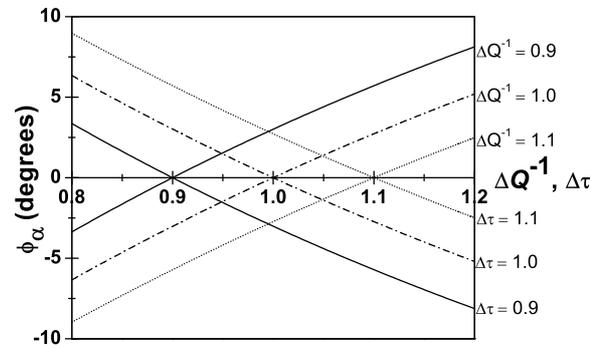}}
\end{center}
\caption{Magnitude (Fig. \ref{Fig:alpha amp}) and phase (Fig. \ref{Fig:alpha phase}) of $\alpha$ for a Newtonian fluid, calculated using Eqs.~\ref{eq:alpha measure mag} and \ref{eq:alpha measure phase}. Each line is plotted at a constant value of either $\Delta Q^{-1}$ or $\Delta\tau$, with the other measured quantity taking the value determined by the $x$-axis. In Eq.~\ref{eq:alpha measure mag} and Fig. \ref{Fig:alpha amp}, the two measured variables are equivalent.} \label{Fig:alpha from measurement}
\end{figure}

Equations~\ref{eq:alpha measure mag} and \ref{eq:alpha measure phase} are plotted in Fig.~\ref{Fig:alpha from measurement}. Figure~\ref{Fig:alpha amp} shows that $\lvert\alpha\rvert$ increases with both $\Delta Q^{-1}$ and $\Delta\tau$, indicating that if mass is decoupled from the surface, the amplitude of fluid oscillation will decrease unless more power is applied. Figure~\ref{Fig:alpha phase} shows that to retain the NSBC value of phase ($\phi_\alpha=0$), $\Delta Q^{-1}$ must decrease if $\Delta\tau$ decreases. Therefore, an oscillation generating relatively low power will tend to produce a value of $\phi_\alpha > 0$ (fluid oscillation leading) unless there is some decoupling of mass from the surface, and therefore reduced inertia. Similarly, if more power is generated, the fluid will lag unless balanced by greater inertia. Considered together, Fig.~\ref{Fig:alpha amp} and Fig.~\ref{Fig:alpha phase} show that a non-zero measurement of either $\Delta Q^{-1}$ or $\Delta\tau$ necessarily indicates a departure from $\lvert\alpha\rvert=1$, or $\phi_\alpha=0$ or both.

\subsubsection{\label{visco}Viscoelasticity}

The introduction of a slip length is mathematically equivalent to considering a viscoelastic fluid. By inspection of Eq.~\ref{eq:Gen interface stress}, and remembering that $\delta\propto\sqrt{\eta}$, we find that instead of using $\alpha$, we can introduce a complex viscosity $\eta^*$ to replace the Newtonian viscosity $\eta$ such that $\alpha\sqrt{\eta}=\sqrt{\eta^*}$. Due to coupling to the decaying wave in the fluid, we have $\sigma\propto\sqrt{\eta^*}\propto \sqrt{\frac{G^*}{i}}$, where $G^*=G'+iG''$ is the complex shear modulus. We find that the storage modulus $G'$ and the loss modulus $G''$ are given by combinations of the inertial and dissipative terms in the oscillation \cite{425}, 

\begin{equation}\label{eq:G'}
\rho G'=(R^2-X^2),
\end{equation}

\noindent and

\begin{equation}\label{eg:G''}
\rho G''=2RX.
\end{equation}

\noindent It is therefore misleading to use the viscoelastic description of \textquoteleft storage' and \textquoteleft loss' moduli for the current situation. It is also misleading to describe a Newtonian fluid as viscoelastic until there is experimental evidence for viscoelastic behaviour rather than some other cause of 
slip. Some previous QCM work based on fully phase-dependent equivalent circuit analyses \cite{477,478, 316} has considered slip in conjunction with a viscoelastic fluid and layering, thereby confusing possible Newtonian slip with other effects; for further effects of layering see the Discussion (below). 
Such an approach may be motivated by layering of a QCM electrode substrate, that is not present when using a torsional oscillator. One previous investigation \cite{313} has noted the equivalence of viscoelasticity and slip in interpretation of slip measurements.

\section{\label{Experimental}Experiment}

\subsection{\label{Details}Experimental details}

To demonstrate investigation of slip using a torsional oscillator, preliminary experiments have been carried out using a 40~kHz oscillator, as described by Fig.~\ref{Fig:TO_Schematic}. The fused quartz specimen rod was uniformly ground using Al$_2$O$_3$ grits to remove surface features of magnitude $>$5~$\mu$m, and polished using cerium oxide powder. Polishing debris was removed by sonication in acetone, isopropanol, 10\% nitric acid and ethanol. Any remaining organic material was removed using \textquoteleft piranha' solution (70~vol\% sulphuric acid, 30~vol\% hydrogen peroxide). The rod was then thoroughly washed in deionised water (0.2~$\mu$m filtered, 18.2~M$\Omega$~cm) and lightly flame-polished to minimize surface roughness. The \textquoteleft slip' case was configured by coating the specimen rod with a hydrophobic fluoropolymer surfactant (RS Components), evenly dispensed from an aerosol can. The base case was retrieved by dissolving the surfactant in isopropanol prior to re-cleaning the rod using \textquoteleft piranha' solution and washing with deionised water and ethanol.  

A pre-optically polished, flame-polished piece of fused quartz has RMS roughness $<1$~nm and peak to trough roughness $<3$~nm, as measured over several widely-distributed areas of size 5~$\mu$m$^2$ using AFM. Previously \cite{418}, Scanning Electron Microscopy (SEM) revealed peak to trough roughness of 2--4~$\mu$m on rods left unpolished following grinding using Al$_2$O$_3$, with no significant alteration of the texture following the addition of the surfactant. However, a flame-polished surface coated with the surfactant has additional texture characterised by features of spacing $<5$~$\mu$m with RMS roughness 24~nm and peak to trough roughness $<300$~nm, observed using optical microscopy and AFM. The advanced static contact angle of deionised water on fused quartz is $44\pm 2$\textdegree. The advanced contact angle on the same surface coated with the surfactant is $118\pm 1$\textdegree. The texture of the coating causes some hysteresis, and the receded contact angle is $113\pm 1$\textdegree. Contact angles were measured using the sessile drop method (CAM 200, KSV Instruments).

Prior to each immersion run, the specimen rod was washed in deionised water and dried in air. For the base case, the uncoated rod was also rinsed with ethanol prior to drying. Deionised water was used as the test fluid in all experiments. The oscillator was contained in an insulated bell jar standing on a plate maintained at 303~$\pm$~1~K and situated on a large mechanically damped pad usually used for a Transmission Electron Microscope.

\subsection{Measurement and calibration}

Measurements from recent work are presented alongside previously reported experiments \cite{418} in Table~\ref{Tab:Results} and explained in this section. The data points for $\Delta Q^{-1}$ and $\Delta\tau$ are each the result of several immersion events in each of the base and slip cases. The specimen rod was immersed by slowly raising a filled vessel to the rod. A value of $Q^{-1}_\eta$ or $\tau_\eta$ for any particular immersion run is the difference between the dry value and the value when the water was in contact with the node of the standing wave, adjusted to account for any background drift. The fluid meniscus was aligned with the vertical mid-point of a rubber o-ring that was fitted half way along the specimen rod. The o-ring, which fixes the position of the node, was located at a point where surface motion is zero or minimised, and is considered rigid during calibration. Each immersion was carried out after the oscillator had been left for at least 90 minutes, so that instrumental drift per second was $<5e^{-5}$ for $V_d$/$V_g$ and $<1e^{-13}$~s for $\tau$, as measured over 10 minute intervals. 

The standard error of the mean measurement was combined with error ascribed to drift to give the total experimental error in $Q^{-1}_\eta$ and $\tau_\eta$ in each case. Uncertainty is dominated by variation between different immersion events, probably due to the extreme thermal and mechanical sensitivity of the instrument. The lesser contribution due to instrumental drift is accounted for by assuming that the mean drift correction is a random error. Earlier experiments \cite{418} have established that uncertainty due to surface tension effects and location of the node is small relative to the standard error in the mean value, because very little damping occurs at or near the node. 

In order to calculate $Q^{-1}_\eta$ from measured voltages, the half width of the oscillation was measured directly following each immersion. Equations~\ref{eq:TO Q^-1} and \ref{eq:TO Q^-1c} were then used to calculate a precise value of $K$. It has been shown \cite{418} that a plot having gradient $K$ is typically linear ($R^2>0.9999$ for all fits) and independent of temperature, the immersion level and the presence of surfactant. The two former variables are held constant during immersions, but the slip case experiments were calibrated separately from the base case in the most recent experiments. The value of $K$ is slightly lower for the slip case, as expected from the first principles analysis \cite{302} in which $K$ is inversely proportional to the oscillator's moment of inertia. The maximum error in $K$, conservatively estimated as the difference between the slip case and the base case, is not significant in comparison with the experimental error of $\alpha$. $K$ changes when the torsional oscillator is mechanically altered, but was constant for each individual shear rate presented in Fig~\ref{Fig:alpha_expt}. 

For the NSBC, it is expected that $Q^{-1}_\eta=2\tau_\eta$/$\tau$ \cite{302}. This relation was followed to within 5\% for the majority of immersions. Even accounting for the possibility of slip, immersion events with $>$10\% deviation clearly indicate a significant instability in the oscillation and such data were discarded.  

Experiments have been performed at shear rate amplitudes $\lvert\dot{\varepsilon}_{r\theta}\rvert$ between 230 and 6840~s$^{-1}$, corresponding to linear displacement amplitudes of 3.2 and 96~nm at the antinode. The linear amplitude of the oscillation is equal to $\sqrt{2}\delta\lvert\dot{\varepsilon}_{r\theta}\rvert$/$\omega$. The shear rate at the surface of the rod is calculated as described by Robinson et al.~\cite{301}. For this calculation, rod diameters and masses were measured using a micrometer screw gauge and a Mettler AE 200 mass balance, respectively. For each strain rate, intrinsic gauge capacitance $C_m$ was measured to better than $\pm 10\%$ by finding the gradient of straight line plots ($R^2>0.999$) of the fractional change in resonant period as a function of total gauge capacitance \cite{301}. The transformer ratio linking the electronic and mechanical parts of the equivalent circuit was calculated using $K$. The alternative calculation (from first principles) incorporates large error relating to the electrode separation angle 2$\Phi$ (Fig.~\ref{Fig:TO_Schematic}, \cite{302}). The final uncertainty in shear rate amplitude is $\sim 15\%$. However, the uncertainty in each value relative to the others depends only on $V_g$, which is precise to better than $1\%$.  

\subsection{\label{Interpretex}Results}

\begin{figure}
\begin{center}
\subfigure[]{\label{Fig:ex alpha amp}\includegraphics[width=8.5cm]{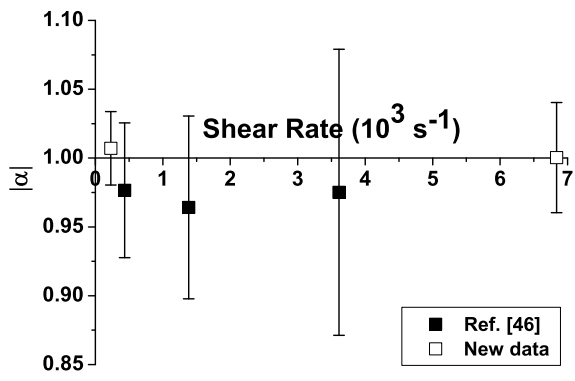}}
\subfigure[]{\label{Fig:ex alpha phase}\includegraphics[width=8.5cm]{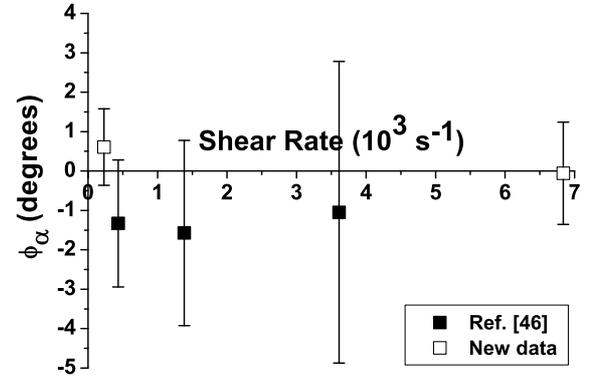}}
\subfigure[]{\label{Fig:ex b amp}\includegraphics[width=8.5cm]{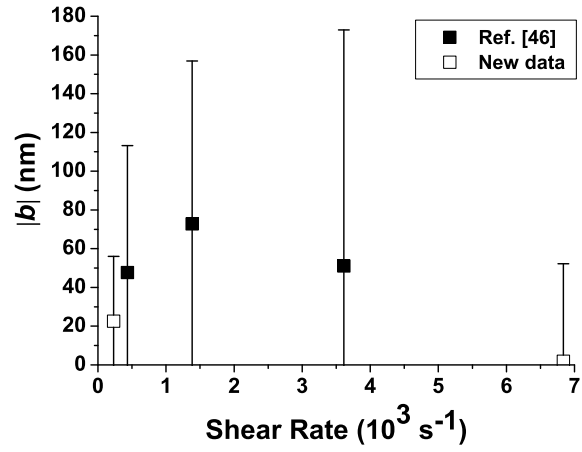}}
\end{center}
\caption{Magnitude (Fig.~\ref{Fig:ex alpha amp}) and phase (Fig.~\ref{Fig:ex alpha phase}) of $\alpha$ and magnitude of $b$ (Fig.~\ref{Fig:ex b amp}) for the experiments in Table~\ref{Tab:Results}. Values of $\lvert b\rvert$ have been updated slightly ($<20$~nm difference in all values) from those presented in Ref.~\cite{418} so that drift is consistently handled. Error bars for the calculated shear rates are omitted for clarity. 
} \label{Fig:alpha_expt}
\end{figure}

The results for the five experiments in Table~\ref{Tab:Results} are plotted as slip parameters in Fig.~\ref{Fig:alpha_expt}. Figures~\ref{Fig:ex alpha amp} and \ref{Fig:ex alpha phase} appear similar because, for all five data points, mean values of $\left(\Delta Q^{-1}-1\right)$ and $\left(\Delta\tau-1\right)$ have the same sign. All data lie within experimental uncertainty of the NSBC. For the three data points from Ref.~\cite{418}, the consistency of mean values ($\lvert\alpha\rvert<1$, $\phi_\alpha<0$) previously led to the conclusion that finite slip had likely been observed. Although the newer results are not directly comparable due to differences in surface preparation, the mean values are now scattered around $\lvert\alpha\rvert=1$, $\phi_\alpha=0$, suggesting that the NSBC is observed with a high degree of accuracy. The most recent experiments have covered a greater range of strain rates and afforded greater precision due to improved stability of the oscillation and better thermal and mechanical isolation. The corresponding mean values of $\lvert b\rvert$ are now less than 30~nm, and the upper limits bounded by experimental uncertainty are less than 60~nm. For each data point, the lower error bar of $\lvert b\vert$ is at zero because the NSBC falls within experimental uncertainty. Because of the phase discontinuity at the NSBC (Fig.~\ref{Fig:alpha vs b}), any value of $\phi_b$ is possible for near-NSBC combinations of the measured variables.

Previously, QCM measurements \cite{306} have obtained values of $b$ of similar precision, although in that case finite slip lengths of magnitudes between 10 and 80~nm were observed on bare as well as a range of hydrophobic surfaces. As discussed previously, the analysis used to derive these values contains inconsistencies. When considering other results in the literature, we emphasize that comparisons are drawn from data sets taken from measurements of very different flows (see Discussion) and that the current data are observed over a well-defined, wide range of shear rates. It is therefore remarkable that the data is consistent with emerging trends from high-precision AFM studies. Namely, slip at a hydrophobic smooth surface occurs on scales $<b=50$~nm, if at all; and surface roughness such as that observed for the slip case in the current work may play a part in suppressing any slip at a hydrophobic surface. 

\section{\label{Discussion}Discussion}

\subsection{\label{Interpretth}Interpretation of slip measurements}

In order to discuss the interpretation of experimental slip measurements, we first clarify the terminology used in the discussion. Lauga et al.'s definitions \cite{335} of \textquoteleft indirect' measurement, \textquoteleft apparent' slip and \textquoteleft effective' slip are used, although we extend the definition of effective slip to include temporal as well as spatial averaging. \textquoteleft True' slip is an oft-used term not defined by Lauga et al. \cite{335}. We define it here as the case when the mean tangential velocity component of fluid is different from that of a solid surface immediately in contact with it. 

Experimental methods are predominantly indirect at present. Even the most \textquoteleft direct' optical methods of slip measurement have been labelled indirect \cite{393}, in the sense that the velocity of tracers rather than the fluid itself is measured. In the case of the oscillator, the measured damping force is not necessarily consistent with the idealised analysis presented in the Theory section. 
In any surface force measurement, the flow profile is inferred rather than explicitly determined. Under these conditions, slip parameters $b$ and $\alpha$ only describe the slip dynamics in full when rate-independent, true slip occurs and the fluid is entirely Newtonian. 

Caution is required when analytically linking postulated mechanisms to apparent or effective slip measurements. For example, it often appears convenient to assume the presence of layering at or near the slip interface \cite{316, 478, 444, 319, 313, 302, 309, 314}. Typically, layers of unknown thickness and interfacial properties only introduce more modelling parameters. As discussed previously, introduced viscoelasticity is indistinguishable from slip. 
In two studies employing $\alpha$ \cite{319, 313}, multiple viscoelastic layers with multiple boundary conditions were modelled and several parameters were fitted. The resulting magnitudes of $\alpha$ (ranging from $\sim 1.8$ to $\sim 5$) were larger than reasonable for a Newtonian fluid. At present, mechanisms of apparent slip are poorly characterized, so empirical measurement of slip parameters remains the most important step.
Newtonian fluids should be considered Newtonian right up to the interface to avoid introducing error.

\subsection{\label{Comp}Comparison with other experimental techniques}

SFA and AFM techniques are currently the most well developed and widely-employed methods of measuring surface slip, and can reach resolution of better than $\pm 5$~nm for $b$ \cite{393, 427}. Optical methods such as the Fluorescent Recovery After Photobleaching (FRAP) \cite{318} and Particle Image Velocimetry (PIV) \cite{440} techniques allow resolution of $\pm$100~nm, limited by diffraction and tracer particle size. Recently, a nano-PIV system has reportedly resolved true slip lengths of less than 100~nm with uncertainty less than $\pm$20~nm \cite{439}. Several groups have measured slip using capillary techniques \cite{303, 335}, but accuracy has been limited by capillary fabrication issues, the difficulty of controlling surface chemistry within a capillary and (in some cases) the effects of electro-osmotic flow. A recently-developed technique derives slip from measurement of thermal motion of confined colloidal particles \cite{441}. A slip length of 18~$\pm$~5~nm has been reported at near-zero shear rate. Torsional oscillator experiments presented in the current work confirmed the NSBC to within $\pm 0.03$ for $\lvert\alpha\rvert$, $\pm 1.5^\degree$ for $\phi_\alpha$ and to within 60~nm of $\lvert b\rvert=0$. We stress that there are no significant limitations on the precision of the oscillator other than control of atmospheric conditions, vibration and electrical noise. Further, the technique offers advantages and points of difference when compared with other techniques.

A particular advantage of the torsional oscillator is characterization of experimental shear rate. The shear rate amplitude 
is easily varied by adjusting the applied gauge voltage and can at least cover the range $10^2$ to $10^4$~s$^{-1}$. A major reason for use of torsional oscillators in rheological experiments is the large and difficult-to-reproduce range of accessible shear rates \cite{419, 431}. In contrast, the variable velocity distribution across a QCM causes analytic complications \cite{313}. In the AFM and SFA techniques \cite{415}, the shear rate varies with respect to surface separation and distance from an axis passing through the centres of the approaching curved surfaces. The shear rate is maximized away from this axis and also depends on the approach or withdrawal rate of the two spheres \cite{410}. 

The second major advantage of the torsional oscillator is versatility. Surface treatments can easily be applied to the specimen rod, either while remote from the oscillator or attached to it. If the specimen rod is detached, the response of the oscillator is easily recalibrated. Electronic voltage and period measurements are captured near-instantaneously, so time-dependent trends can be measured. The oscillator is perfectly suited to measurement of in-situ changes of surface chemistry or other parameters, and experimental precision would be significantly enhanced for such experiments. The oscillator can readily be used at variable temperatures and even variable pressures. Although it is best suited to differential measurements (similarly to optical techniques), the zero-slip response can be determined by careful calibration of all the parameters described in the original equivalent circuit analysis \cite{301,302}. In contrast, use of QCMs and other torsional oscillator designs \cite{424, 425, 431} for slip measurement presents difficulties with regard to surface modification and incorporating piezoelectric equations into an equivalent circuit analysis. 

An important distinction between damping-oscillation methods and other techniques is that the shear damping force is measured. SFA and AFM methods measure drainage forces caused by pressure in the flow around curved surfaces; viscous forces are not significant. Another difference is that, when using the torsional oscillator, slip occurs over a relatively large fluid-solid interface, so measurement of molecular-scale effects is limited by surface homogeneity. However, measurement of effective slip will give more accuracy and precision than a value derived from a small system. It is not true that measurements of effective slip must take place on a scale $L\sim b$ in order to have a measurable impact \cite{335}, because any fundamental slip mechanism should be consistent over the whole interface. 
Regardless, we note that the linear displacement amplitude for the torsional oscillator is of comparable magnitude to the slip length, so molecular-scale interactions are certainly probed using this device.

\section{Conclusion}

In this paper, we have drawn particular attention to the occurrence of slip at an oscillating interface. The ultrasonic composite torsional oscillator is a versatile instrument for measurement of viscous damping forces, and therefore slip. In order to fully characterize this device for slip measurements, we have revisited fundamental aspects of the slip analysis. The analysis has then been applied at a generalised oscillating surface, followed by the surface of the oscillator's specimen rod in particular. Preliminary experiments at shear rates between 230 and 6800~s$^{-1}$ have probed slip at a hydrophobic surface of RMS roughness 24~nm, obtaining resolution better than $\pm 0.03$ for $\lvert\alpha\rvert$, $\pm 1.5^\degree$ for $\phi_\alpha$ and within 60~nm of $\lvert b\rvert=0$. 

Navier's slip parameter, while retaining the units of length, is best suited to interpretation as a surface force effect. Measurement of surface forces has played an important role in developing the understanding of slip to date, while applications are likely to utilise the lubricating effect at a slipping surface. The actual velocity discontinuity at an oscillating interface, as encapsulated by the parameter $\alpha$ for oscillating surfaces, is being directly observed with increasing precision using optical methods. The link between a velocity discontinuity and a surface force necessarily utilizes the spatial gradient of an analytical flow profile, unless both types of experiment can be performed on the same system.

Ongoing experiments will aim to significantly develop the stock of published slip measurements at oscillating surfaces. Particular experimental topics will be the influence of shear rate, hydrophobicity and surface texture, while the development and in-situ characterisation of surfaces or fluids with switchable properties will also be a priority. 

\section*{Acknowledgements}

Thanks are due to Dr. Shen Chong for assistance with setting up the apparatus described here and used previously \cite{418}.

\makeatletter
\def\@seccntformat#1{\csname the#1\endcsname:\quad}
\makeatother

\renewcommand{\theequation}{A\arabic{equation}}
\renewcommand{\thesection}{APPENDIX \Alph{section}}
\setcounter{equation}{0}  
\setcounter{section}{0}

\nohang\section{\label{AppA}Derivation of Slip-Adjusted Torque Applied to a Specimen Rod}

In this section, the damping torque on a composite torsional ultrasonic oscillator rod, as illustrated in Fig. \ref{Fig:TO_Schematic}, is calculated. We follow Robinson and Smedley \cite{302} and incorporate the case in which there is finite slip, as introduced in the Theory section. Using the cylindrical polar co-ordinate system defined in Fig.~\ref{Fig:TO_Schematic}, the velocity in the azimuthal ($\theta$) direction is given by

\begin{equation}\label{eq:TO NSBC rodspeed}
u_\theta\left(r,z,t\right) = u_{s,max}\left(t\right)\cos\left(\frac{2\pi z}{\lambda}\right),
\end{equation}

\noindent where the subscript \textquoteleft max' refers to the value at $r=a$ and at the antinode of the standing wave, which has wavelength $\lambda$.

There is a velocity gradient in the fluid in both the $r$- and $z$-directions. Viscous forces in the fluid exert azimuthal shear stresses on the rod, which dampen the torsional oscillation. Forces due to the $z$-gradient are neglected (Appendix~B). The total damping torque is derived by integrating the tangential stress (Eq.~\ref{eq:Gen interface stress}) along the immersed, curved length of the rod $l$. It is assumed that the fluid surface coincides with a node and the free end is an antinode, so that $\sin\left(\frac{2\pi l}{\lambda}\right)=1$. The torque is 

\begin{equation}\label{eq:TO NSBC Trod}
\begin{split}
T_{rod}\left(t\right)&= \int_{0}^{l}\sigma_{\theta r}\left(t\right) 2\pi a^2 dz\\
&=\frac{2\sqrt{2i}}{\delta}\pi a^2\eta
\alpha u_{s,max}\left(t\right)\int_{0}^{l}\cos\Bigl(\frac{2\pi z}{\lambda}\Bigr)dz\\
&=\frac{\sqrt{2i}}{\delta}a^2\lambda\eta\alpha u_{s,max}\left(t\right).
\end{split}
\end{equation}

The torque on the oscillating flat end of a rod is derived similarly. The surface velocity at radius $r$ from the cylindrical axis is given by

\begin{equation}\label{eq:TO NSBC end solution}
u_\theta\left(r,z=0,t\right)=\frac{r}{a}u_{s,max}\left(t\right),
\end{equation}

and the torque is then integrated over the surface of the free end, so that

\begin{equation}\label{eq:TO NSBC Tend}
\begin{split}
T_{end}\left(t\right)&= \int_{0}^{a}\sigma_{\theta z}\left(t\right)2\pi r^2dr\\&=\frac{2\sqrt{2i}}{\delta a}\pi\eta
\alpha u_{s,max}\left(t\right)\int_{0}^{a}r^3dr\\ &=\frac{\sqrt{2i}}{\delta}\frac{\pi
a^3}{2}\eta\alpha u_{s,max}\left(t\right),
\end{split}
\end{equation}

Neglecting edge effects between the flat end and the curved surfaces, the total torque for $h$ immersed flat surfaces is 

\begin{equation}\label{eqApp:TO NSBC T}
T\left(t\right)=a^2(\lambda+\frac{1}{2}h\pi a)\frac{\sqrt{2i}}{\delta}\eta\alpha u_{s,max}\left(t\right).
\end{equation}

\renewcommand{\theequation}{B\arabic{equation}}

\setcounter{equation}{0}   

\nohang\section{\label{App no z}Viscous Gradient in the $z$-direction}

The velocity flow field around the torsional rod has gradients in both the $r$- (radial) and $z$- (axial) directions. The azimuthal stress due to these gradients are given by

\begin{equation}\label{eq:TO NSBC no z}
\begin{split}
\sigma_{\theta r} &= \eta\frac{\partial u_\theta\left(r,z,t\right)}{\partial r}\vline_{r=a}\\
&\sim\frac{-\sqrt{2i}}{\delta}u_{\theta}\left(a,z,t\right)
\end{split}
\end{equation}
and
\begin{equation}\label{eq:TO NSBC no z2}
\begin{split}
\sigma_{\theta z}\left(t\right)&= \eta\frac{\partial u_\theta\left(r,z,t\right)}{\partial z}\vline_{r=a}\\
&\sim\frac{2\pi}{\lambda}u_\theta\left(a,z,t\right).
\end{split}
\end{equation}

\noindent For a 40 kHz torsional oscillator in water, the ratio of radial to axial stresses is $\sim10^4$.

\clearpage

\clearpage

\clearpage

\clearpage

\clearpage

\clearpage

\clearpage

\begin{table}
\caption{Details of experiments previously presented (\cite{418}, labelled \textquoteleft A') and new experiments (labelled \textquoteleft B').
}
\label{Tab:Results}
\begin{center}
\begin{tabular}{ccccccc}\hline\hline
Experiment&$\lvert\dot{\varepsilon}_{r\theta}\rvert$&Linear Amplitude&&$K$&$\Delta Q^{-1}$&$\Delta\tau$\\
&(10$^3$~s$^{-1}$)&(nm)&&(10$^{-5}$)&&\\\hline
A-1&$0.43\pm 0.06$&$6.1\pm 0.9$&&0.93&$0.99\pm 0.01$&$0.95\pm 0.05$\\
A-2&$1.4\pm 0.2$&$19\pm 3$&&0.87&$0.99\pm 0.01$&$0.94\pm 0.07$\\
A-3&$3.6\pm 0.5$&$51\pm 7$&&0.36&$0.99\pm 0.02$&$0.96\pm 0.10$\\\hline
B-1&$0.23\pm 0.03$&$3.2\pm 0.5$&(base)&3.60&$1.00\pm 0.01$&$1.02 \pm 0.03$\\
&&&(slip)&3.59&&\\
B-2&$6.8\pm 1.0$&$96\pm 14$&(base)&3.57&$1.00\pm 0.04$&$1.00\pm 0.01$\\
&&&(slip)&3.57&&\\
\hline\hline
\end{tabular}
\end{center}
\end{table}


\clearpage

\begin{thebibliography}{58}
\expandafter\ifx\csname natexlab\endcsname\relax\def\natexlab#1{#1}\fi
\expandafter\ifx\csname bibnamefont\endcsname\relax
  \def\bibnamefont#1{#1}\fi
\expandafter\ifx\csname bibfnamefont\endcsname\relax
  \def\bibfnamefont#1{#1}\fi
\expandafter\ifx\csname citenamefont\endcsname\relax
  \def\citenamefont#1{#1}\fi
\expandafter\ifx\csname url\endcsname\relax
  \def\url#1{\texttt{#1}}\fi
\expandafter\ifx\csname urlprefix\endcsname\relax\def\urlprefix{URL }\fi
\providecommand{\bibinfo}[2]{#2}
\providecommand{\eprint}[2][]{\url{#2}}

\bibitem[{\citenamefont{Neto et~al.}(2005)\citenamefont{Neto, Evans,
  Bonaccurso, Butt, and Craig}}]{303}
\bibinfo{author}{\bibfnamefont{C.}~\bibnamefont{Neto}},
  \bibinfo{author}{\bibfnamefont{D.~R.} \bibnamefont{Evans}},
  \bibinfo{author}{\bibfnamefont{E.}~\bibnamefont{Bonaccurso}},
  \bibinfo{author}{\bibfnamefont{H.-J.} \bibnamefont{Butt}}, \bibnamefont{and}
  \bibinfo{author}{\bibfnamefont{V.~S.~J.} \bibnamefont{Craig}},
  \bibinfo{journal}{Rep. Prog. Phys.} \textbf{\bibinfo{volume}{68}},
  \bibinfo{pages}{2859} (\bibinfo{year}{2005}).

\bibitem[{\citenamefont{Lauga et~al.}(2005)\citenamefont{Lauga, Brenner, and
  Stone}}]{335}
\bibinfo{author}{\bibfnamefont{E.}~\bibnamefont{Lauga}},
  \bibinfo{author}{\bibfnamefont{M.~P.} \bibnamefont{Brenner}},
  \bibnamefont{and} \bibinfo{author}{\bibfnamefont{H.~A.} \bibnamefont{Stone}},
  \emph{\bibinfo{title}{Handbook of Experimental Fluid Dynamics}}
  (\bibinfo{publisher}{Springer}, \bibinfo{address}{New York},
  \bibinfo{year}{2005}), Chap. \bibinfo{chapter}{15}.

\bibitem[{\citenamefont{Bocquet and Barrat}(2007)\citenamefont{Bocquet and Barrat}}]{502}
\bibinfo{author}{\bibfnamefont{L.}~\bibnamefont{Bocquet}} \bibnamefont{and}
  \bibinfo{author}{\bibfnamefont{J.-L.} \bibnamefont{Barrat}},
  \bibinfo{journal}{Soft Matter} \textbf{\bibinfo{volume}{3}},
  \bibinfo{pages}{685} (\bibinfo{year}{2007}).

\bibitem[{\citenamefont{Squires and Quake}(2005)}]{321}
\bibinfo{author}{\bibfnamefont{T.~M.} \bibnamefont{Squires}} \bibnamefont{and}
  \bibinfo{author}{\bibfnamefont{S.~R.} \bibnamefont{Quake}},
  \bibinfo{journal}{Rev. Mod. Phys.} \textbf{\bibinfo{volume}{77}},
  \bibinfo{pages}{977} (\bibinfo{year}{2005}).

\bibitem[{\citenamefont{Vinogradova}(1999)}]{325}
\bibinfo{author}{\bibfnamefont{O.~I.} \bibnamefont{Vinogradova}},
  \bibinfo{journal}{Int. J. Miner. Process.} \textbf{\bibinfo{volume}{56}},
  \bibinfo{pages}{31} (\bibinfo{year}{1999}).

\bibitem[{\citenamefont{Granick et~al.}(2003)\citenamefont{Granick, Zhu, and
  Lee}}]{403}
\bibinfo{author}{\bibfnamefont{S.}~\bibnamefont{Granick}},
  \bibinfo{author}{\bibfnamefont{Y.}~\bibnamefont{Zhu}}, \bibnamefont{and}
  \bibinfo{author}{\bibfnamefont{H.}~\bibnamefont{Lee}}, \bibinfo{journal}{Nat.
  Mater.} \textbf{\bibinfo{volume}{2}}, \bibinfo{pages}{221}
  (\bibinfo{year}{2003}).

\bibitem[{\citenamefont{Navier}(1827)}]{400}
\bibinfo{author}{\bibfnamefont{C.~L. M.~H.} \bibnamefont{Navier}},
  \bibinfo{journal}{Mem. Acad. Sci. Inst. Fr.} \textbf{\bibinfo{volume}{6}},
  \bibinfo{pages}{839} (\bibinfo{year}{1827}).

\bibitem[{\citenamefont{Vinogradova}(1995)}]{305}
\bibinfo{author}{\bibfnamefont{O.~I.} \bibnamefont{Vinogradova}},
  \bibinfo{journal}{Langmuir} \textbf{\bibinfo{volume}{11}},
  \bibinfo{pages}{2213} (\bibinfo{year}{1995}).

\bibitem[{\citenamefont{Cottin-Bizonne
  et~al.}(2003)\citenamefont{Cottin-Bizonne, Barrat, Bocquet, and
  Charlaix}}]{404}
\bibinfo{author}{\bibfnamefont{C.}~\bibnamefont{Cottin-Bizonne}},
  \bibinfo{author}{\bibfnamefont{J.-L.} \bibnamefont{Barrat}},
  \bibinfo{author}{\bibfnamefont{L.}~\bibnamefont{Bocquet}}, \bibnamefont{and}
  \bibinfo{author}{\bibfnamefont{E.}~\bibnamefont{Charlaix}},
  \bibinfo{journal}{Nat. Mater.} \textbf{\bibinfo{volume}{2}},
  \bibinfo{pages}{237} (\bibinfo{year}{2003}).

\bibitem[{\citenamefont{Lauga and Stone}(2003)}]{329}
\bibinfo{author}{\bibfnamefont{E.}~\bibnamefont{Lauga}} \bibnamefont{and}
  \bibinfo{author}{\bibfnamefont{H.~A.} \bibnamefont{Stone}},
  \bibinfo{journal}{J. Fluid Mech.} \textbf{\bibinfo{volume}{489}},
  \bibinfo{pages}{55} (\bibinfo{year}{2003}).

\bibitem[{\citenamefont{Brochard and de~Gennes}(1992)}]{411}
\bibinfo{author}{\bibfnamefont{F.}~\bibnamefont{Brochard}} \bibnamefont{and}
  \bibinfo{author}{\bibfnamefont{P.~G.} \bibnamefont{de~Gennes}},
  \bibinfo{journal}{Langmuir} \textbf{\bibinfo{volume}{8}},
  \bibinfo{pages}{3033} (\bibinfo{year}{1992}).

\bibitem[{\citenamefont{Horn et~al.}(2000)\citenamefont{Horn, Vinogradova,
  Mackay, and Phan-Thien}}]{336}
\bibinfo{author}{\bibfnamefont{R.~G.} \bibnamefont{Horn}},
  \bibinfo{author}{\bibfnamefont{O.~I.} \bibnamefont{Vinogradova}},
  \bibinfo{author}{\bibfnamefont{M.~E.} \bibnamefont{Mackay}},
  \bibnamefont{and}
  \bibinfo{author}{\bibfnamefont{N.}~\bibnamefont{Phan-Thien}},
  \bibinfo{journal}{J. Chem. Phys.} \textbf{\bibinfo{volume}{112}},
  \bibinfo{pages}{6424} (\bibinfo{year}{2000}).

\bibitem[{\citenamefont{Leger et~al.}(1997)\citenamefont{Leger, Hervet, Massey,
  and Durliat}}]{433}
\bibinfo{author}{\bibfnamefont{L.}~\bibnamefont{Leger}},
  \bibinfo{author}{\bibfnamefont{H.}~\bibnamefont{Hervet}},
  \bibinfo{author}{\bibfnamefont{G.}~\bibnamefont{Massey}}, \bibnamefont{and}
  \bibinfo{author}{\bibfnamefont{E.}~\bibnamefont{Durliat}},
  \bibinfo{journal}{J. Phys. - Condens. Mat.} \textbf{\bibinfo{volume}{9}},
  \bibinfo{pages}{7719} (\bibinfo{year}{1997}).

\bibitem[{\citenamefont{Ybert et~al.}(unpublished)\citenamefont{Ybert,
  Barentin, Cottin-Bizonne, Joseph, and Bocquet}}]{430}
\bibinfo{author}{\bibfnamefont{C.}~\bibnamefont{Ybert}},
  \bibinfo{author}{\bibfnamefont{C.}~\bibnamefont{Barentin}},
  \bibinfo{author}{\bibfnamefont{C.}~\bibnamefont{Cottin-Bizonne}},
  \bibinfo{author}{\bibfnamefont{P.}~\bibnamefont{Joseph}}, \bibnamefont{and}
  \bibinfo{author}{\bibfnamefont{L.}~\bibnamefont{Bocquet}}
  (\bibinfo{year}{unpublished}).

\bibitem[{\citenamefont{Joseph et~al.}(2006)\citenamefont{Joseph,
  Cottin-Bizonne, Benoit, Ybert, Journet, Tabeling, and Bocquet}}]{437}
\bibinfo{author}{\bibfnamefont{P.}~\bibnamefont{Joseph}},
  \bibinfo{author}{\bibfnamefont{C.}~\bibnamefont{Cottin-Bizonne}},
  \bibinfo{author}{\bibfnamefont{J.~M.} \bibnamefont{Benoit}},
  \bibinfo{author}{\bibfnamefont{C.}~\bibnamefont{Ybert}},
  \bibinfo{author}{\bibfnamefont{C.}~\bibnamefont{Journet}},
  \bibinfo{author}{\bibfnamefont{P.}~\bibnamefont{Tabeling}}, \bibnamefont{and}
  \bibinfo{author}{\bibfnamefont{L.}~\bibnamefont{Bocquet}},
  \bibinfo{journal}{Phys. Rev. Lett.} \textbf{\bibinfo{volume}{97}},
  \bibinfo{pages}{156104} (\bibinfo{year}{2006}).

\bibitem[{\citenamefont{Choi and Kim}(2006)}]{320}
\bibinfo{author}{\bibfnamefont{C.-H.} \bibnamefont{Choi}} \bibnamefont{and}
  \bibinfo{author}{\bibfnamefont{C.-J.} \bibnamefont{Kim}},
  \bibinfo{journal}{Phys. Rev. Lett.} \textbf{\bibinfo{volume}{96}},
  \bibinfo{pages}{066001} (\bibinfo{year}{2006}).

\bibitem[{\citenamefont{Choi et~al.}(2006)\citenamefont{Choi, Ulmanella, Kim,
  Ho, and Kim}}]{436}
\bibinfo{author}{\bibfnamefont{C.~H.} \bibnamefont{Choi}},
  \bibinfo{author}{\bibfnamefont{U.}~\bibnamefont{Ulmanella}},
  \bibinfo{author}{\bibfnamefont{J.}~\bibnamefont{Kim}},
  \bibinfo{author}{\bibfnamefont{C.~M.} \bibnamefont{Ho}}, \bibnamefont{and}
  \bibinfo{author}{\bibfnamefont{C.~J.} \bibnamefont{Kim}},
  \bibinfo{journal}{Phys. Fluids} \textbf{\bibinfo{volume}{18}},
  \bibinfo{pages}{087105} (\bibinfo{year}{2006}).

\bibitem[{\citenamefont{Ou and Rothstein}(2005)}]{434}
\bibinfo{author}{\bibfnamefont{J.}~\bibnamefont{Ou}} \bibnamefont{and}
  \bibinfo{author}{\bibfnamefont{J.~P.} \bibnamefont{Rothstein}},
  \bibinfo{journal}{Phys. Fluids} \textbf{\bibinfo{volume}{17}},
  \bibinfo{pages}{103606} (\bibinfo{year}{2005}).

\bibitem[{\citenamefont{Gogte et~al.}(2005)\citenamefont{Gogte, Vorobieff,
  Truesdell, Mammoli, van Swol, Shah, and Brinker}}]{435}
\bibinfo{author}{\bibfnamefont{S.}~\bibnamefont{Gogte}},
  \bibinfo{author}{\bibfnamefont{P.}~\bibnamefont{Vorobieff}},
  \bibinfo{author}{\bibfnamefont{R.}~\bibnamefont{Truesdell}},
  \bibinfo{author}{\bibfnamefont{A.}~\bibnamefont{Mammoli}},
  \bibinfo{author}{\bibfnamefont{F.}~\bibnamefont{van Swol}},
  \bibinfo{author}{\bibfnamefont{P.}~\bibnamefont{Shah}}, \bibnamefont{and}
  \bibinfo{author}{\bibfnamefont{C.~J.} \bibnamefont{Brinker}},
  \bibinfo{journal}{Phys. Fluids} \textbf{\bibinfo{volume}{17}},
  \bibinfo{pages}{051701} (\bibinfo{year}{2005}).

\bibitem[{\citenamefont{Chan and Horn}(1985)}]{415}
\bibinfo{author}{\bibfnamefont{D.~Y.~C.} \bibnamefont{Chan}} \bibnamefont{and}
  \bibinfo{author}{\bibfnamefont{R.~G.} \bibnamefont{Horn}},
  \bibinfo{journal}{J. Chem. Phys.} \textbf{\bibinfo{volume}{83}},
  \bibinfo{pages}{5311} (\bibinfo{year}{1985}).

\bibitem[{\citenamefont{Craig et~al.}(2001)\citenamefont{Craig, Neto, and
  Williams}}]{304}
\bibinfo{author}{\bibfnamefont{V.~S.~J.} \bibnamefont{Craig}},
  \bibinfo{author}{\bibfnamefont{C.}~\bibnamefont{Neto}}, \bibnamefont{and}
  \bibinfo{author}{\bibfnamefont{D.~R.~M.} \bibnamefont{Williams}},
  \bibinfo{journal}{Phys. Rev. Lett.} \textbf{\bibinfo{volume}{87}},
  \bibinfo{pages}{054504} (\bibinfo{year}{2001}).

\bibitem[{\citenamefont{Bonaccurso et~al.}(2003)\citenamefont{Bonaccurso, Butt,
  and Craig}}]{324}
\bibinfo{author}{\bibfnamefont{E.}~\bibnamefont{Bonaccurso}},
  \bibinfo{author}{\bibfnamefont{H.-J.} \bibnamefont{Butt}}, \bibnamefont{and}
  \bibinfo{author}{\bibfnamefont{V.~S.~J.} \bibnamefont{Craig}},
  \bibinfo{journal}{Phys. Rev. Lett.} \textbf{\bibinfo{volume}{90}},
  \bibinfo{pages}{144501} (\bibinfo{year}{2003}).

\bibitem[{\citenamefont{Cottin-Bizonne
  et~al.}(2005)\citenamefont{Cottin-Bizonne, Cross, Steinberger, and
  Charlaix}}]{393}
\bibinfo{author}{\bibfnamefont{C.}~\bibnamefont{Cottin-Bizonne}},
  \bibinfo{author}{\bibfnamefont{B.}~\bibnamefont{Cross}},
  \bibinfo{author}{\bibfnamefont{A.}~\bibnamefont{Steinberger}},
  \bibnamefont{and} \bibinfo{author}{\bibfnamefont{E.}~\bibnamefont{Charlaix}},
  \bibinfo{journal}{Phys. Rev. Lett.} \textbf{\bibinfo{volume}{94}},
  \bibinfo{pages}{056102} (\bibinfo{year}{2005}).

\bibitem[{\citenamefont{Cross et~al.}(to be published)\citenamefont{Cross,
  Steinberger, Cottin-Bizonne, Rieu, and Charlaix}}]{427}
\bibinfo{author}{\bibfnamefont{B.}~\bibnamefont{Cross}},
  \bibinfo{author}{\bibfnamefont{A.}~\bibnamefont{Steinberger}},
  \bibinfo{author}{\bibfnamefont{C.}~\bibnamefont{Cottin-Bizonne}},
  \bibinfo{author}{\bibfnamefont{J.~P.} \bibnamefont{Rieu}}, \bibnamefont{and}
  \bibinfo{author}{\bibfnamefont{E.}~\bibnamefont{Charlaix}},
  \bibinfo{journal}{Europhys. Lett.}  (\bibinfo{year}{to be published}).

\bibitem[{\citenamefont{Pit et~al.}(2000)\citenamefont{Pit, Hervet, and
  Leger}}]{318}
\bibinfo{author}{\bibfnamefont{R.}~\bibnamefont{Pit}},
  \bibinfo{author}{\bibfnamefont{H.}~\bibnamefont{Hervet}}, \bibnamefont{and}
  \bibinfo{author}{\bibfnamefont{L.}~\bibnamefont{Leger}},
  \bibinfo{journal}{Phys. Rev. Lett.} \textbf{\bibinfo{volume}{85}},
  \bibinfo{pages}{980} (\bibinfo{year}{2000}).

\bibitem[{\citenamefont{Bonaccurso et~al.}(2002)\citenamefont{Bonaccurso,
  Kappl, and Butt}}]{334}
\bibinfo{author}{\bibfnamefont{E.}~\bibnamefont{Bonaccurso}},
  \bibinfo{author}{\bibfnamefont{M.}~\bibnamefont{Kappl}}, \bibnamefont{and}
  \bibinfo{author}{\bibfnamefont{H.-J.} \bibnamefont{Butt}},
  \bibinfo{journal}{Phys. Rev. Lett.} \textbf{\bibinfo{volume}{88}},
  \bibinfo{pages}{076103} (\bibinfo{year}{2002}).

\bibitem[{\citenamefont{Huang and Breuer}(2007)}]{439}
\bibinfo{author}{\bibfnamefont{P.}~\bibnamefont{Huang}} \bibnamefont{and}
  \bibinfo{author}{\bibfnamefont{K.~S.} \bibnamefont{Breuer}},
  \bibinfo{journal}{Phys. Fluids} \textbf{\bibinfo{volume}{19}},
  \bibinfo{pages}{028104} (\bibinfo{year}{2007}).

\bibitem[{\citenamefont{L.~Joly and Bocquet}(2006)}]{441}
\bibinfo{author}{\bibfnamefont{C.~Y.} \bibnamefont{L.~Joly}} \bibnamefont{and}
  \bibinfo{author}{\bibfnamefont{L.}~\bibnamefont{Bocquet}},
  \bibinfo{journal}{Phys. Rev. Lett.} \textbf{\bibinfo{volume}{96}},
  \bibinfo{pages}{046101} (\bibinfo{year}{2006}).

\bibitem[{\citenamefont{Vinogradova and Yakubov}(2006)}]{413}
\bibinfo{author}{\bibfnamefont{O.~I.} \bibnamefont{Vinogradova}}
  \bibnamefont{and} \bibinfo{author}{\bibfnamefont{G.~E.}
  \bibnamefont{Yakubov}}, \bibinfo{journal}{Phys. Rev. E}
  \textbf{\bibinfo{volume}{73}}, \bibinfo{pages}{045302(R)}
  (\bibinfo{year}{2006}).

\bibitem[{\citenamefont{Honig and Ducker}(2007)}]{410}
\bibinfo{author}{\bibfnamefont{C.~D.~F.} \bibnamefont{Honig}} \bibnamefont{and}
  \bibinfo{author}{\bibfnamefont{W.~A.} \bibnamefont{Ducker}},
  \bibinfo{journal}{Phys. Rev. Lett.} \textbf{\bibinfo{volume}{98}},
  \bibinfo{pages}{028305} (\bibinfo{year}{2007}).

\bibitem[{\citenamefont{Richardson}(1973)}]{483}
\bibinfo{author}{\bibfnamefont{S.}~\bibnamefont{Richardson}},
  \bibinfo{journal}{J. Fluid Mech.} \textbf{\bibinfo{volume}{59}},
  \bibinfo{pages}{707} (\bibinfo{year}{1973}).

\bibitem[{\citenamefont{Kanazawa and II}(1985)}]{331}
\bibinfo{author}{\bibfnamefont{K.~K.} \bibnamefont{Kanazawa}} \bibnamefont{and}
  \bibinfo{author}{\bibfnamefont{J.~G.~G.} \bibnamefont{II}},
  \bibinfo{journal}{Anal. Chim. Acta} \textbf{\bibinfo{volume}{175}},
  \bibinfo{pages}{99} (\bibinfo{year}{1985}).

\bibitem[{\citenamefont{Ellis and Hayward}(2003)}]{306}
\bibinfo{author}{\bibfnamefont{J.~S.} \bibnamefont{Ellis}} \bibnamefont{and}
  \bibinfo{author}{\bibfnamefont{G.~L.} \bibnamefont{Hayward}},
  \bibinfo{journal}{J. Appl. Phys.} \textbf{\bibinfo{volume}{94}},
  \bibinfo{pages}{7856} (\bibinfo{year}{2003}).

\bibitem[{\citenamefont{Hayward and Thompson}(1998)}]{313}
\bibinfo{author}{\bibfnamefont{G.~L.} \bibnamefont{Hayward}} \bibnamefont{and}
  \bibinfo{author}{\bibfnamefont{M.}~\bibnamefont{Thompson}},
  \bibinfo{journal}{J. Appl. Phys.} \textbf{\bibinfo{volume}{83}},
  \bibinfo{pages}{2194} (\bibinfo{year}{1998}).

\bibitem[{\citenamefont{Ferrante et~al.}(1994)\citenamefont{Ferrante, Kipling,
  and Thompson}}]{319}
\bibinfo{author}{\bibfnamefont{F.}~\bibnamefont{Ferrante}},
  \bibinfo{author}{\bibfnamefont{A.~L.} \bibnamefont{Kipling}},
  \bibnamefont{and} \bibinfo{author}{\bibfnamefont{M.}~\bibnamefont{Thompson}},
  \bibinfo{journal}{J. Appl. Phys.} \textbf{\bibinfo{volume}{76}},
  \bibinfo{pages}{3448} (\bibinfo{year}{1994}).

\bibitem[{\citenamefont{Salomaki and Kankare}(2004)}]{316}
\bibinfo{author}{\bibfnamefont{M.}~\bibnamefont{Salomaki}} \bibnamefont{and}
  \bibinfo{author}{\bibfnamefont{J.}~\bibnamefont{Kankare}},
  \bibinfo{journal}{Langmuir} \textbf{\bibinfo{volume}{20}},
  \bibinfo{pages}{7794} (\bibinfo{year}{2004}).

\bibitem[{\citenamefont{Ellis et~al.}(2003)\citenamefont{Ellis, McHale,
  Hayward, and Thompson}}]{308}
\bibinfo{author}{\bibfnamefont{J.~S.} \bibnamefont{Ellis}},
  \bibinfo{author}{\bibfnamefont{G.}~\bibnamefont{McHale}},
  \bibinfo{author}{\bibfnamefont{G.~L.} \bibnamefont{Hayward}},
  \bibnamefont{and} \bibinfo{author}{\bibfnamefont{M.}~\bibnamefont{Thompson}},
  \bibinfo{journal}{J. Appl. Phys.} \textbf{\bibinfo{volume}{94}},
  \bibinfo{pages}{6201} (\bibinfo{year}{2003}).

\bibitem[{\citenamefont{McHale and Newton}(2004)}]{309}
\bibinfo{author}{\bibfnamefont{G.}~\bibnamefont{McHale}} \bibnamefont{and}
  \bibinfo{author}{\bibfnamefont{M.~I.} \bibnamefont{Newton}},
  \bibinfo{journal}{J. Appl. Phys.} \textbf{\bibinfo{volume}{95}},
  \bibinfo{pages}{373} (\bibinfo{year}{2004}).

\bibitem[{\citenamefont{McHale et~al.}(2000)\citenamefont{McHale, Lucklum,
  Newton, and Cowen}}]{314}
\bibinfo{author}{\bibfnamefont{G.}~\bibnamefont{McHale}},
  \bibinfo{author}{\bibfnamefont{R.}~\bibnamefont{Lucklum}},
  \bibinfo{author}{\bibfnamefont{M.~I.} \bibnamefont{Newton}},
  \bibnamefont{and} \bibinfo{author}{\bibfnamefont{J.~A.} \bibnamefont{Cowen}},
  \bibinfo{journal}{J. Appl. Phys.} \textbf{\bibinfo{volume}{88}},
  \bibinfo{pages}{7304} (\bibinfo{year}{2000}).

\bibitem[{\citenamefont{Mason}(1949)}]{432}
\bibinfo{author}{\bibfnamefont{W.~P.} \bibnamefont{Mason}},
  \bibinfo{journal}{Trans. Am. Soc. Mech. Eng.} \textbf{\bibinfo{volume}{69}},
  \bibinfo{pages}{359} (\bibinfo{year}{1949}).

\bibitem[{\citenamefont{Bergenholtz et~al.}(1998)\citenamefont{Bergenholtz,
  Willenbacher, Wagner, Morrison, van~den Ende, and Mellema}}]{425}
\bibinfo{author}{\bibfnamefont{J.}~\bibnamefont{Bergenholtz}},
  \bibinfo{author}{\bibfnamefont{N.}~\bibnamefont{Willenbacher}},
  \bibinfo{author}{\bibfnamefont{N.~J.} \bibnamefont{Wagner}},
  \bibinfo{author}{\bibfnamefont{B.}~\bibnamefont{Morrison}},
  \bibinfo{author}{\bibfnamefont{D.}~\bibnamefont{van~den Ende}},
  \bibnamefont{and} \bibinfo{author}{\bibfnamefont{J.}~\bibnamefont{Mellema}},
  \bibinfo{journal}{J. Colloid Interf. Sci.} \textbf{\bibinfo{volume}{202}},
  \bibinfo{pages}{430} (\bibinfo{year}{1998}).

\bibitem[{\citenamefont{Weiss et~al.}(1999)\citenamefont{Weiss, Ballauff, and
  Willenbacher}}]{426}
\bibinfo{author}{\bibfnamefont{A.}~\bibnamefont{Weiss}},
  \bibinfo{author}{\bibfnamefont{M.}~\bibnamefont{Ballauff}}, \bibnamefont{and}
  \bibinfo{author}{\bibfnamefont{N.}~\bibnamefont{Willenbacher}},
  \bibinfo{journal}{J. Colloid Interf. Sci.} \textbf{\bibinfo{volume}{216}},
  \bibinfo{pages}{185} (\bibinfo{year}{1999}).

\bibitem[{\citenamefont{Fritz et~al.}(2003)\citenamefont{Fritz, Pechhold,
  Willenbacher, and Wagner}}]{431}
\bibinfo{author}{\bibfnamefont{G.}~\bibnamefont{Fritz}},
  \bibinfo{author}{\bibfnamefont{W.}~\bibnamefont{Pechhold}},
  \bibinfo{author}{\bibfnamefont{N.}~\bibnamefont{Willenbacher}},
  \bibnamefont{and} \bibinfo{author}{\bibfnamefont{N.~J.}
  \bibnamefont{Wagner}}, \bibinfo{journal}{J. Rheol.}
  \textbf{\bibinfo{volume}{47}}, \bibinfo{pages}{303} (\bibinfo{year}{2003}).

\bibitem[{\citenamefont{Kelarakis et~al.}(2006)\citenamefont{Kelarakis,
  Crassous, and Ballauff}}]{419}
\bibinfo{author}{\bibfnamefont{A.}~\bibnamefont{Kelarakis}},
  \bibinfo{author}{\bibfnamefont{J.~J.} \bibnamefont{Crassous}},
  \bibnamefont{and} \bibinfo{author}{\bibfnamefont{M.}~\bibnamefont{Ballauff}},
  \bibinfo{journal}{Langmuir} \textbf{\bibinfo{volume}{22}},
  \bibinfo{pages}{6814} (\bibinfo{year}{2006}).

\bibitem[{\citenamefont{Robinson et~al.}(1974)\citenamefont{Robinson,
  Carpenter, and Tallon}}]{301}
\bibinfo{author}{\bibfnamefont{W.~H.} \bibnamefont{Robinson}},
  \bibinfo{author}{\bibfnamefont{S.~H.} \bibnamefont{Carpenter}},
  \bibnamefont{and} \bibinfo{author}{\bibfnamefont{J.~L.}
  \bibnamefont{Tallon}}, \bibinfo{journal}{J. Appl. Phys.}
  \textbf{\bibinfo{volume}{45}}, \bibinfo{pages}{1975} (\bibinfo{year}{1974}).

\bibitem[{\citenamefont{Robinson and Smedley}(1978)}]{302}
\bibinfo{author}{\bibfnamefont{W.~H.} \bibnamefont{Robinson}} \bibnamefont{and}
  \bibinfo{author}{\bibfnamefont{S.~I.} \bibnamefont{Smedley}},
  \bibinfo{journal}{J. Appl. Phys.} \textbf{\bibinfo{volume}{49}},
  \bibinfo{pages}{1070} (\bibinfo{year}{1978}).

\bibitem[{\citenamefont{Willmott and Tallon}(2007)}]{418}
\bibinfo{author}{\bibfnamefont{G.~R.} \bibnamefont{Willmott}} \bibnamefont{and}
  \bibinfo{author}{\bibfnamefont{J.~L.} \bibnamefont{Tallon}},
  \bibinfo{journal}{Curr. Appl. Phys.}
  (\bibinfo{year}{to be published}).

\bibitem[{\citenamefont{Thompson and Troian}(1997)}]{482}
\bibinfo{author}{\bibfnamefont{P.~A.} \bibnamefont{Thompson}} \bibnamefont{and}
  \bibinfo{author}{\bibfnamefont{S.~M.} \bibnamefont{Troian}},
  \bibinfo{journal}{Nature} \textbf{\bibinfo{volume}{389}},
  \bibinfo{pages}{360} (\bibinfo{year}{1997}).

\bibitem[{\citenamefont{Zhu and Granick}(2001)}]{327}
\bibinfo{author}{\bibfnamefont{Y.}~\bibnamefont{Zhu}} \bibnamefont{and}
  \bibinfo{author}{\bibfnamefont{S.}~\bibnamefont{Granick}},
  \bibinfo{journal}{Phys. Rev. Lett.} \textbf{\bibinfo{volume}{87}},
  \bibinfo{pages}{096105} (\bibinfo{year}{2001}).

\bibitem[{\citenamefont{de~Gennes}(2002)}]{444}
\bibinfo{author}{\bibfnamefont{P.~G.} \bibnamefont{de~Gennes}},
  \bibinfo{journal}{Langmuir} \textbf{\bibinfo{volume}{18}},
  \bibinfo{pages}{3413} (\bibinfo{year}{2002}).

\bibitem[{\citenamefont{Lu et~al.}(2004)\citenamefont{Lu, Lee, and Lim}}]{312}
\bibinfo{author}{\bibfnamefont{F.}~\bibnamefont{Lu}},
  \bibinfo{author}{\bibfnamefont{H.~P.} \bibnamefont{Lee}}, \bibnamefont{and}
  \bibinfo{author}{\bibfnamefont{S.~P.} \bibnamefont{Lim}},
  \bibinfo{journal}{J. Phys. D Appl. Phys.} \textbf{\bibinfo{volume}{37}},
  \bibinfo{pages}{898} (\bibinfo{year}{2004}).

\bibitem[{\citenamefont{Du et~al.}(2004)\citenamefont{Du, Goubaidoulline, and
  Johannsmann}}]{307}
\bibinfo{author}{\bibfnamefont{B.}~\bibnamefont{Du}},
  \bibinfo{author}{\bibfnamefont{I.}~\bibnamefont{Goubaidoulline}},
  \bibnamefont{and}
  \bibinfo{author}{\bibfnamefont{D.}~\bibnamefont{Johannsmann}},
  \bibinfo{journal}{Langmuir} \textbf{\bibinfo{volume}{20}},
  \bibinfo{pages}{10617} (\bibinfo{year}{2004}).

\bibitem[{\citenamefont{Daikhin et~al.}(2000)\citenamefont{Daikhin, Gileadi,
  Tsionsky, Urbakh, and Zilberman}}]{315}
\bibinfo{author}{\bibfnamefont{L.}~\bibnamefont{Daikhin}},
  \bibinfo{author}{\bibfnamefont{E.}~\bibnamefont{Gileadi}},
  \bibinfo{author}{\bibfnamefont{V.}~\bibnamefont{Tsionsky}},
  \bibinfo{author}{\bibfnamefont{M.}~\bibnamefont{Urbakh}}, \bibnamefont{and}
  \bibinfo{author}{\bibfnamefont{G.}~\bibnamefont{Zilberman}},
  \bibinfo{journal}{Electrochim. Acta} \textbf{\bibinfo{volume}{45}},
  \bibinfo{pages}{3615} (\bibinfo{year}{2000}).

\bibitem[{\citenamefont{Rodahl and Kasemo}(1996)}]{395}
\bibinfo{author}{\bibfnamefont{M.}~\bibnamefont{Rodahl}} \bibnamefont{and}
  \bibinfo{author}{\bibfnamefont{B.}~\bibnamefont{Kasemo}},
  \bibinfo{journal}{Sensor. Actuat. A - Phys.} \textbf{\bibinfo{volume}{54}},
  \bibinfo{pages}{448} (\bibinfo{year}{1996}).

\bibitem[{\citenamefont{Hendy et~al.}(2005)\citenamefont{Hendy, Jasperse, and
  Burnell}}]{317}
\bibinfo{author}{\bibfnamefont{S.~C.} \bibnamefont{Hendy}},
  \bibinfo{author}{\bibfnamefont{M.}~\bibnamefont{Jasperse}}, \bibnamefont{and}
  \bibinfo{author}{\bibfnamefont{J.}~\bibnamefont{Burnell}},
  \bibinfo{journal}{Phys. Rev. E} \textbf{\bibinfo{volume}{72}},
  \bibinfo{pages}{016303} (\bibinfo{year}{2005}).

\bibitem[{\citenamefont{Hafer and Laesecke}(2003)}]{424}
\bibinfo{author}{\bibfnamefont{R.~F.} \bibnamefont{Hafer}} \bibnamefont{and}
  \bibinfo{author}{\bibfnamefont{A.}~\bibnamefont{Laesecke}},
  \bibinfo{journal}{Meas. Sci. Technol.} \textbf{\bibinfo{volume}{14}},
  \bibinfo{pages}{663} (\bibinfo{year}{2003}).

\bibitem[{\citenamefont{Lea and Fozooni}(1985)}]{477}
\bibinfo{author}{\bibfnamefont{M.~J.} \bibnamefont{Lea}} \bibnamefont{and}
  \bibinfo{author}{\bibfnamefont{P.}~\bibnamefont{Fozooni}},
  \bibinfo{journal}{Ultrasonics} \textbf{\bibinfo{volume}{23}},
  \bibinfo{pages}{133} (\bibinfo{year}{1985}).

\bibitem[{\citenamefont{Yang et~al.}(1993)\citenamefont{Yang, Thompson, and
  Duncan-Hewitt}}]{478}
\bibinfo{author}{\bibfnamefont{M.}~\bibnamefont{Yang}},
  \bibinfo{author}{\bibfnamefont{M.}~\bibnamefont{Thompson}}, \bibnamefont{and}
  \bibinfo{author}{\bibfnamefont{W.~C.} \bibnamefont{Duncan-Hewitt}},
  \bibinfo{journal}{Langmuir} \textbf{\bibinfo{volume}{9}},
  \bibinfo{pages}{802} (\bibinfo{year}{1993}).

\bibitem[{\citenamefont{Joseph and Tabeling}(2005)}]{440}
\bibinfo{author}{\bibfnamefont{P.}~\bibnamefont{Joseph}} \bibnamefont{and}
  \bibinfo{author}{\bibfnamefont{P.}~\bibnamefont{Tabeling}},
  \bibinfo{journal}{Phys. Rev. E} \textbf{\bibinfo{volume}{71}},
  \bibinfo{pages}{035303(R)} (\bibinfo{year}{2005}).

\end{thebibliography}
\end{document}